\documentclass[10pt,a4paper]{article}
\usepackage[latin1]{inputenc}
\usepackage{amsmath, amsfonts, amssymb, amsthm, mathtools}
\usepackage{bm}
\usepackage{graphicx}
\usepackage{geometry}
\usepackage[english]{babel} 
\usepackage{booktabs} 
\usepackage{abstract} 
\usepackage{authblk}
\usepackage{array}
\usepackage{todonotes}
\usepackage{subcaption}
\usepackage{rotating}
\usepackage{pdflscape}
\usepackage[ruled, linesnumbered]{algorithm2e}

\usepackage{hyperref} 
\usepackage{apacite}

\usepackage{xcolor}
\definecolor{myorange}{RGB}{201,53,56}
\hypersetup{colorlinks=True,citecolor=myorange}

\theoremstyle{definition}
\newtheorem{theorem}{Theorem}[section]

\newtheorem{proposition}[theorem]{Proposition}

\theoremstyle{definition}

\newtheorem{definition}[theorem]{Definition}
\newtheorem{example}[theorem]{Example}

\theoremstyle{remark}
\newtheorem{remark}[theorem]{Remark}

\newcommand{\varphinn}{\varphi_\textrm{NN}}
\newcommand{\varphitilde}{\tilde{\varphi}}
\newcommand{\dd}{\ensuremath{\mathrm{d}}}

\newcommand*\colvec[3][]{
	\begin{pmatrix}\ifx\relax#1\relax\else#1\\\fi#2\\#3\end{pmatrix}
}
\DeclareMathOperator*{\argmin}{arg\,min}
\newcommand{\norm}[1]{\left\lVert#1\right\rVert}

\begin{document}
	\title{Deep calibration of rough stochastic volatility models}
	
	\author[1]{Christian Bayer}
	\author[1,2]{Benjamin Stemper}

	\affil[1]{Weierstrass Institute for Applied Analysis and Stochastics, Berlin, Germany}
	\affil[2]{Department of Mathematics, Technical University Berlin, Berlin, Germany}
		\maketitle
		\begin{abstract}
		\noindent
		Sparked by \citeA{ALV07, Fuk11, Fuk17, GJR18}, so-called \emph{rough} stochastic volatility models such as the rough Bergomi model by \citeA{BFG16} constitute the latest evolution in option price modeling. Unlike standard bivariate diffusion models such as \citeA{Hes93}, these non-Markovian models with fractional volatility drivers  allow to parsimoniously recover key stylized facts of market implied volatility surfaces such as the exploding power-law behaviour of the at-the-money volatility skew as time to maturity goes to zero. Standard model calibration routines rely on the repetitive evaluation of the map from model parameters to Black-Scholes implied volatility, rendering calibration of many (rough) stochastic volatility models prohibitively expensive since there the map can often only be approximated by costly Monte Carlo (MC) simulations \cite{BLP17, MP18, BFG16, HJM17}. As a remedy, we propose to combine a standard Levenberg-Marquardt calibration routine with neural network regression, replacing expensive MC simulations with cheap forward runs of a neural network trained to approximate the implied volatility map. Numerical experiments confirm the high accuracy and speed of our approach.
		\vspace{0.5cm}
	\end{abstract}


\section{Introduction}
\begin{figure*}
	\centering
	\includegraphics[width=0.7\linewidth]{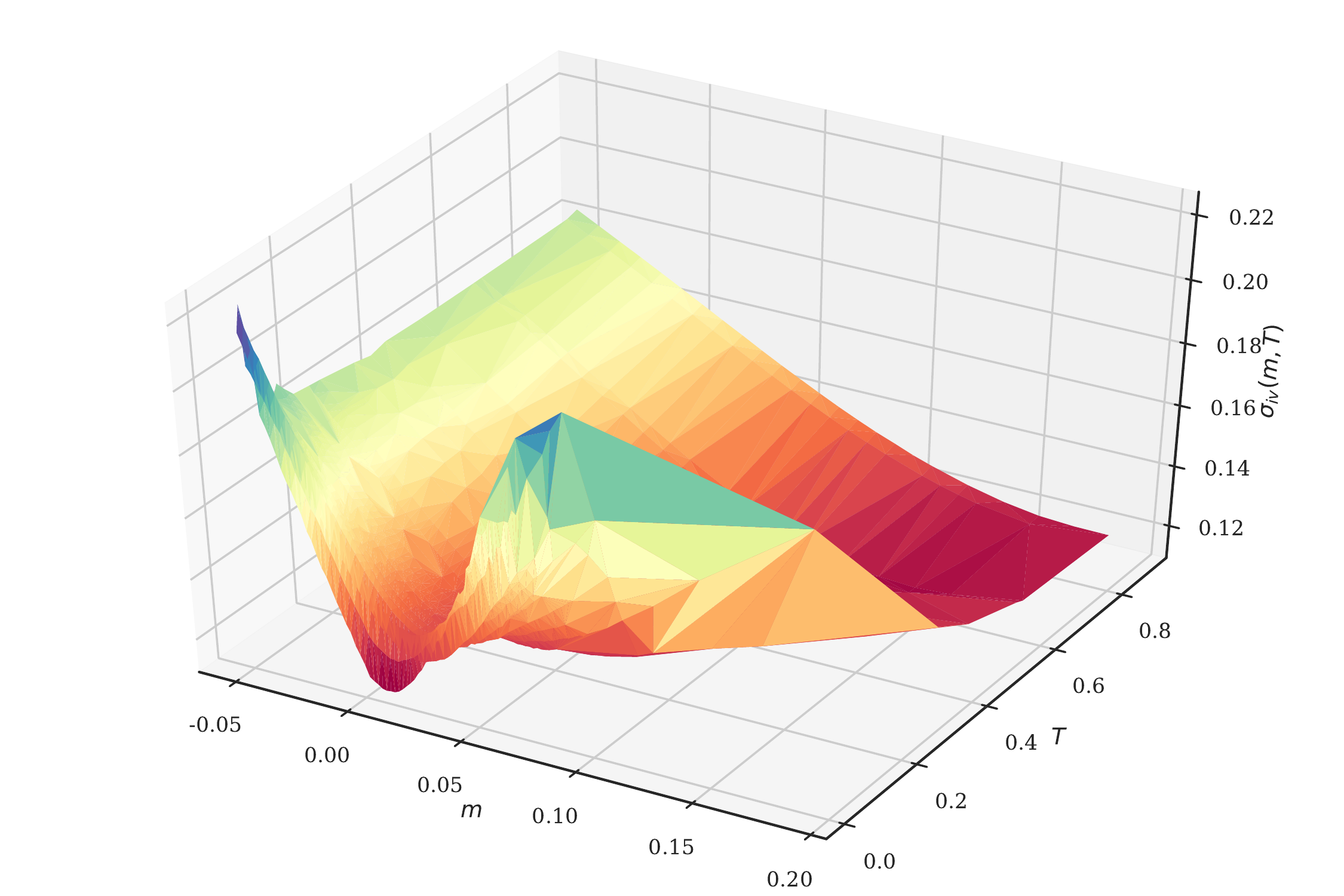}
	\caption{\textbf{SPX Market Implied Volatility surface on 15th February 2018.} IVs have been inverted from SPX Weekly European plain vanilla Call Mid prices and the interpolation is a (non-arbitrage-free) Delaunay triangulation. Axes denote log-moneyness $m = \log(K/S_0)$ for strike $K$ and spot $S_0$, time to maturity $T$ in years and market implied volatility $\sigma_\textrm{iv}(m,T)$.}
	\label{fig:spxiv150218}
\end{figure*}

Almost half a century after its publication, the option pricing model by \citeA{BS73} remains one of the most popular analytical frameworks for pricing and hedging European options in financial markets. A part of its success stems from the availability of explicit and hence instantaneously computable closed formulas for both theoretical option prices and option price sensitivities to input parameters (\emph{Greeks}), albeit at the expense of assuming that \emph{volatility} -- the standard deviation of log returns of the underlying asset price -- is deterministic and constant. Still, in financial practice, the Black-Scholes model is often considered a sophisticated transform between option prices and Black-Scholes (BS) \emph{implied volatility (IV)} $\sigma_\textrm{iv}$ where the latter is defined as the constant volatility input needed in the BS formula to match a given (market) price. It is a well-known fact that in empirical IV surfaces obtained by transforming market prices of European options to IVs, it can be observed that IVs vary across moneyness and maturities, exhibiting well-known smiles and at-the-money (ATM) skews and thereby contradicting the flat surface predicted by Black-Scholes (Figure \ref{fig:spxiv150218}). 
In particular, \citeA{BFG16} report empirical at-the-money volatility skews of the form 
\begin{align}
\label{eq: atmskewpower}
	\left|\frac{\partial}{\partial m}\sigma_\textrm{iv}(m,T)\right|  \sim T^{-0.4}, \quad T \to 0
\end{align}
for log moneyness $m$ and time to maturity $T$.

While plain vanilla European Call and Put options often show enough liquidity to be marked-to-market, pricing and hedging path-dependent options (so-called \emph{Exotics}) necessitates an option pricing model that prices European options \emph{consistently} with respect to observed market IVs across moneyness and maturities. In other words, it should parsimoniously capture stylized facts of empirical IV surfaces. To address the shortcomings of Black-Scholes and incorporate the stochastic nature of volatility itself, popular bivariate diffusion models such as SABR \cite{HKLW02} or the ones by \citeA{Hes93} or \citeA{HW90} have been developed to capture \emph{some} important stylized facts. However, according to \citeA{Gat11}, diffusive stochastic volatility models in general fail to recover the exploding power-law nature \eqref{eq: atmskewpower} of the volatility skew as time to maturity goes to 0 and instead predict a constant behaviour.

Sparked by the seminal work of \citeA{ALV07, Fuk11, Fuk17, GJR18}, we have since seen a shift from classical diffusive modeling towards so-called rough stochastic volatility models. They may be defined as a class of \emph{continuous-path} stochastic volatility models where the instantaneous volatility is driven by a stochastic process with H\"older regularity smaller than Brownian Motion, typically modeled by a fractional Brownian Motion with Hurst parameter $H<\frac{1}{2}$ \cite{MN68}. The evidence for this paradigm shift is by now overwhelming, both under the physical measure where time series analysis suggests that log realized volatility has H\"older regularity in the order of $\approx 0.1$ \cite{BLP16, GJR18} and also under the pricing measure where the empirically observed power-law behaviour of the volatility skew near zero may be reproduced in the model \cite{ALV07, Fuk11, Fuk17, BFG16, BFGHS17}. Serious computational and mathematical challenges arise from the non-Markovianity of fractional Brownian motion, effectively forcing researchers to resort to asymptotic expansions \cite{FZ17, BFGHS17} in limiting regimes or (variance-reduced) Monte Carlo schemes \cite{BFG16, BLP17, BFGMS17, HJM17, MP18} to compute fair option prices.

Model calibration is the optimization procedure of finding model parameters such that the IV surface induced by the model best approximates a given market IV surface in an appropriate metric. In the absence of an analytical solution, it is standard practice to solve the arising weighted non-linear least squares problem using iterative optimizers such as Levenberg-Marquardt (LM) \cite{Lev44, Mar63}. However, these optimizers rely on the repetitive evaluation of the function $\varphi$ from the space of model \& option parameters (and external market information) to model BS implied volatility. If each such evaluation involves a time-- and/or memory--intensive operation such as a Monte Carlo simulation in the case of \emph{rough Bergomi} \cite{BFG16} or other (rough) stochastic volatility models, this makes efficient calibration prohibitively expensive.

Made possible by theoretical advancements as well as the widespread availability of cheap, high performance computing hardware, \emph{Machine Learning} has seen a tremendous rise in popularity among academics and practitioners in recent years. Breakthroughs such as (super-) human level performance in image classification \cite{KSH12, SZ14, SLJ15} or playing the ancient Chinese board game \emph{Go} \cite{SSS17} may all be attributed to the advent of \emph{Deep Learning} \cite{GBC16}. Fundamentally, its success stems from the capability of multi-layered artificial neural networks to closely approximate functions $f$ only implicitly available through input-output pairs $\{(x_i, f(x_i))\}_{i=1}^N$, so-called \emph{labeled data}.

The fundamental idea of this paper is to leverage this capability by training a fully-connected neural network on specifically tailored, synthetically generated training data to learn a map $\varphinn$ approximating the true implied volatility map $\varphi$. 
\begin{remark}
	In a related but different approach, \citeA{Her17} proposes to use a neural network to learn the complete calibration routine -- denoted $\Psi$ in our notation in \eqref{eq:calibration_objective} -- taking market data as inputs and returning calibrated model parameters directly. He demonstrates numerically the prowess of his approach by calibrating the popular short rate model of \citeA{HW90} to market data. 
\end{remark}
Both generating the synthetic data set as well as the actual neural network training are expensive in time and computing resource requirements, yet they only have to be performed a single time. Trained networks may then be quickly and efficiently saved, moved and deployed. The benefit of this novel approach is twofold: First, evaluations of $\varphinn$ amount to cheap and almost instantaneous forward runs of a pre-trained network. Second, automatic differentiation of $\varphinn$ with respect to the model parameters returns fast and accurate approximations of the Jacobians needed for the LM calibration routine. Used together, they allow for the efficient calibration of \emph{any} (rough) stochastic volatility model including \emph{rough Bergomi}. 

To demonstrate the practical benefits of our approach numerically, we apply our machinery to \citeA{Hes93} and \emph{rough Bergomi} \cite{BFG16} as representatives of classical and (rough) stochastic volatility models respectively. Speed-wise, no \emph{systematic} comparison is made between the proposed neural network based approach and existing methods, yet with about 40ms per evaluation, our approach is at least competitive with existing Heston pricing methods and beats state-of-the-art rough Bergomi pricing schemes by magnitudes. Also, in both experiments, $\varphinn$ exhibits small relative errors across the highly-liquid parts of the IV surface, recovering characteristic IV smiles and ATM IV skews. To quantify the uncertainty about model parameter estimates obtained by calibrating with $\varphinn$, we infer model parameters in a Bayesian spirit from (i) a synthetically generated IV surface and (ii) SPX market IV data. In both experiments, a simple (weighted) Bayesian nonlinear regression returns a (joint) posterior distribution over model parameters that (1) correctly identifies sensible model parameter regions and (2) places its peak at or close to the true (in the case of the synthetic IV) or previously reported \cite{BFG16} (in the case of the SPX surface) model parameter values. Both experiments thus confirm the idea that $\varphinn$ is sufficiently accurate for calibration.

This paper is organized as follows. In Section \ref{sec:background}, we set the scene, introduce notation and revisit some important machinery that lies at the core of our proposed calibration scheme. In Section \ref{sec:calibration_general}, we state the model calibration objective and introduce \emph{deep calibration}, our approach of combining the established Levenberg-Marquardt calibration algorithm with neural network regression to enable the efficient calibration of (rough) stochastic volatility models. In Section \ref{sec:nn_training}, we outline practical intricacies of our approach, ranging from considerations related to generating synthetic, tailored \emph{labeled data} for training, validation and testing to tricks of the trade when training neural networks and performing hyperparameter optimization. Finally, in Section \ref{sec:numerical_results}, we collect the results of our numerical experiments.

\section{Background}
\label{sec:background}
We now set the scene and introduce notation. Throughout the paper, we shall be working on a filtered probability space $\left(\Omega, \mathcal{F} , \{\mathcal{F}_t\}_{t\geq 0}, \mathbb{P}\right)$ satisfying the \emph{usual conditions} and supporting two (or more) independent Brownian motions under the pricing measure $\mathbb{P}$. We consider a finite time horizon $T < \infty$ and assume the asset price process $S=(S_t)_{t\in [0,T]}$ has been without loss of generality normalized such that spot $S_0=1$ and risk-free rate $r=0$. We define \emph{moneyness} $M:=K/S_0$ and \emph{log moneyness} $m := \log(M) = \log(K) $.

\subsection{Construction of a model IV surface}
\label{subsec:model_iv_surface}

The concept of an \emph{implied volatility surface} is an important idea and tool central to the theory of modern option pricing. In the introduction, we saw how such a surface arises from market prices of liquid European Call options on the S\&P 500 Index \emph{SPX} (cf. Figure~\ref{fig:spxiv150218}). We now formalize the construction of such a surface from model prices. In a first step, we define the pricing function that maps model \& option parameters (and possibly external market information) to the fair price of a European option at time $t=0$.
\begin{definition}[Pricing map]
	\label{def:general_pricing_map}
	Consider a (rough) stochastic volatility (market) model for an asset $S$ with model parameters $\mu  \in \mathcal{M} \subseteq \mathbb{R}^{m}$ and possibly incorporated market information 	$\xi \in \mathcal{E} \subseteq \mathbb{R}^{k}$. The fair price of a European Call option at time $t=0$ is then given by 
	\begin{align*}
	\mathbb{E}\left[S_T(\mu, \xi) - M\right]^+
	\end{align*}
	where $(M,T) \in \Theta \subseteq \mathbb{R}^{2}$ denote moneyness and time to maturity respectively. Letting
	\begin{align}
	\label{def:input_space_pricing}
		\mathcal{I} := \{(\mu, \xi) \times (M,T) \mid \mu \in \mathcal{M}, \xi \in \mathcal{E}, (M,T)^T \in \Theta\} \subseteq \mathbb{R}^{m +k+2}
	\end{align}
	be the pricing input space, we then define the pricing map $P_0: \mathcal{I} \to \mathbb{R}_+$ by
	\begin{align}
	\label{eq: pricing_map_p}
		(\mu, \xi) \times (M,T) \mapsto \mathbb{E}\left[S_T(\mu, \xi) - M\right]^+.
	\end{align}
\end{definition}

\begin{example}
	\label{ex:dc_roughbergomi}
	In the rough Bergomi model by \citeA {BFG16}, the dynamics for the asset price process $S$ and the instantaneous variance process $v=(v_t)_{t \in [0,T]}$ are given by
	\begin{align*}
	\frac{\dd S_t}{S_t} &= \sqrt{v_t} \dd \left(\rho W_t + \sqrt{1-\rho^2}W^{\bot}_t \right) \\
	v_t &= \xi_0 (t) \exp\left(\eta W_t^H  - \frac{1}{2} \eta^2 t^{2H}\right), \quad t \in [0,T].
	\intertext{Here, $\left(W, W^{\bot}\right)=\left(W_t, W_t^{\bot}\right)_{t \in [0,T]}$ are two independent Brownian motions and $\rho \in (-1,1) $ is a constant correlation parameter introducing the \emph{leverage effect} -- the empirically observed anti correlation between stock and volatility movements -- at the driving noise level. The parameter $\eta > 0$ denotes volatility of variance and $\xi_0(t): \mathbb{R}_+ \to \mathbb{R}_+$ given by $	\xi_0(t) = \mathbb{E}(v_t), t\in [0,T]$ 
is a so-called \emph{forward variance curve} which may be recovered from market information \cite{BFG16}. Moreover, $W^H$ is a \emph{Riemann-Liouville} fractional Brownian motion given by}
	W_t^H &= \sqrt{2H}\int_0^t (t-s)^{H-\frac{1}{2}} \dd W_s, \quad t \in [0,T]
	\end{align*}
	with Hurst parameter $H \in (0,1)$. By Kolmogorov, sample paths of $W^H$ are locally almost surely H-$\varepsilon$ H\"older for $\varepsilon > 0$. With respect to Definition \ref{def:general_pricing_map}, hence $\mu=(H, \eta, \rho)$ and $\xi = \xi_0$.
	
\end{example}
\begin{example} 
	\label{ex:dc_heston}
	In the Heston model \cite{Hes93}, with independent Brownian motions $W$ and $ W^\bot$ and model parameters $\rho, \eta$ defined as in Example \ref{ex:dc_roughbergomi}, the dynamics of the asset price $S$ and the instantaneous variance process $v = (v_t)_{t \in [0,T]} $ starting from spot variance $v_0 > 0$ follow
	\begin{align*}
	\frac{\dd S_t}{S_t} &= \sqrt{v_t} \dd \left(\rho W_t + \sqrt{1-\rho^2}W^{\bot}_t \right)\\
	\dd v_t &= \lambda (\bar{v}- v_t) \dd t + \eta \sqrt{v_t} \dd W_t, \quad t \in [0,T].
	\end{align*}
	Here, $\bar{v} > 0$ is the long-run average variance and $\lambda > 0$ is the speed of mean reversion. Feller's condition $2 \lambda \bar{v} >\eta^2$ ensures that $v_t 
	> 0$ for $t\geq 0$. In this model, we thus have  $\mu = (\lambda, \bar{v}, v_0,\rho, \eta)$  and no market information is incorporated into the model.
\end{example}
Let $\textrm{BS}(M, T, \sigma)$ denote the Black-Scholes price of a European Call with moneyness $M$, time to maturity $T$ and assumed constant volatility $\sigma$ of the underlying and let $Q(M,T)$ be the corresponding market price. The BS implied volatility $\sigma_\textrm{iv}(M,T)$ corresponding to $Q(M,T)$ satisfies
	\begin{align*}
		Q(M, T) - \textrm{BS}(M, T, \sigma_\textrm{iv}(M,T)) \overset{!}{=} 0.
	\end{align*}
and the map $(M,T) \mapsto \sigma_\textrm{iv}(M,T)$ is called a \emph{volatility surface}.

\begin{definition}[IV map]
	Let $\mu, \xi, M, T$ be defined as in Definition \ref{def:general_pricing_map}. The Black-Scholes IV $\sigma_\textrm{iv}(\mu, \xi, M, T)$ corresponding to the theoretical model price $P_0\left(\mu, \xi, M,T\right)$ satisfies
	\begin{align}
	P_0\left(\mu, \xi, M,T\right) - \textrm{BS}(M, T, \sigma_\textrm{iv}(\mu, \xi, M, T)) \overset{!}{=} 0.
	\end{align}
	The function $\varphi: \mathcal{I} \to \mathbb{R}_+$ given by
	\begin{align}
	\label{eq:iv_map}
		(\mu, \xi, M,T) \mapsto \sigma_\textrm{iv}(\mu, \xi, M, T)
	\end{align}
	is what we call the \emph{implied volatility map}.	
\end{definition}

\subsection{Regression with neural networks}
\label{subsec:fcnn}

\begin{figure}
	\centering	
		\includegraphics[width=.5\textwidth]{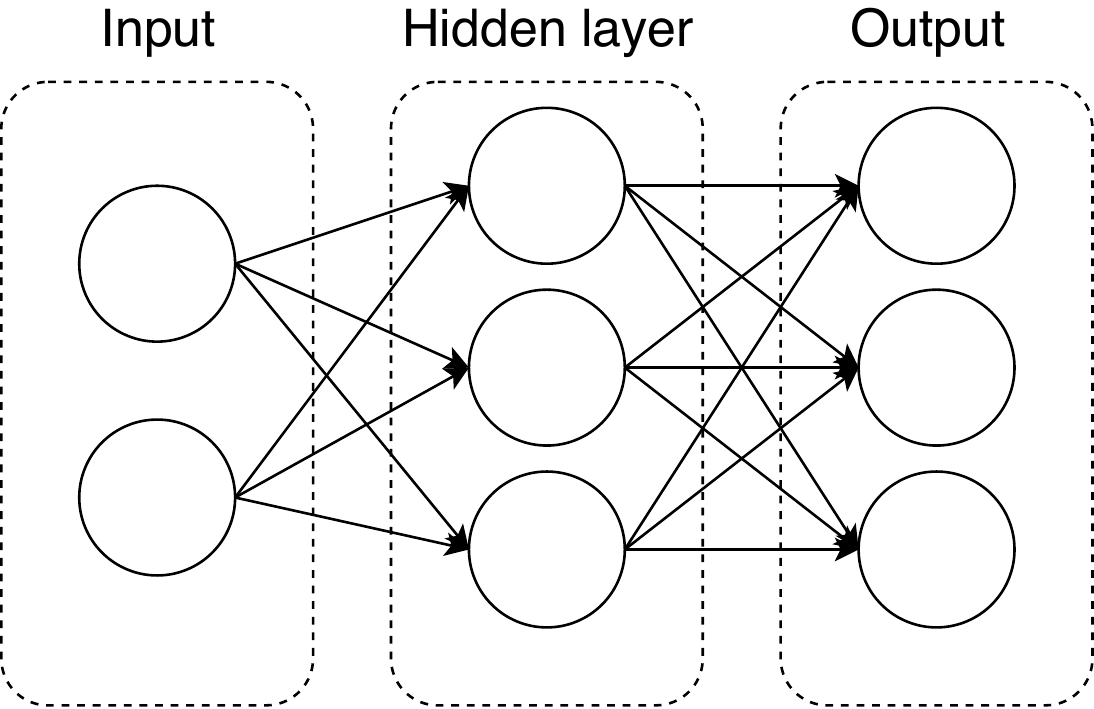}
		\caption{\textbf{Schematic of a fully-connected neural network (FCNN).} Depicted FCNN has a single \emph{hidden layer} consisting of three neurons and may learn to represent a subset of general functions $f: \mathbb{R}^2 \to \mathbb{R}^3$. In the directed acyclic graph above, vertices denote nodes and directed edges describe the flow of information from one node to the next. If number of hidden layers higher than one (typically dozens or hundreds of layers), a neural network is considered \emph{deep}.}
	\label{fig:fc_nn}
\end{figure}

Given a data set $\mathcal{D}=\left\{(x_i, y_i) : x_i \in \mathbb{R}^d, y_i \in \mathbb{R}\right\}_{i=1}^n$ of variables $x_i$ and corresponding scalar, continuous response variables $y_i$, the statistical procedure of estimating the relationship between these variables is commonly called \emph{regression analysis}. Here, we will introduce neural networks and outline their prowess as a regression tool.

The atomic building block of every neural network is a \emph{node}, a functional that performs a weighted sum of its (multi-dimensional) inputs, adds a bias term and then composes the linearity with a scalar non-linear function $\alpha: \mathbb{R} \to \mathbb{R}$ that is identical across the network. Formally, for some input $x \in \mathbb{R}^d, d \in \mathbb{N},$ the output of an individual node is given by
\begin{align*}
	y = \alpha\left(w^T x + b\right) \in \mathbb{R}
\end{align*}
where $w \in \mathbb{R}^d$ and $b\in \mathbb{R}$ are individual \emph{weight} and \emph{bias} terms. An \emph{artificial neural network} is then a collection of many such nodes, grouped into non-overlapping sets called \emph{layers} together with a rule of how the information flows between the layers. 

Over the years different architectural styles have been developed to suit the specific needs of different domains such as speech, text or vision. Arguably the simplest neural network topology not adapted to any particular domain is that of a \emph{fully-connected neural network} (FCNN). An FCNN consists of sequentially ordered so-called \emph{dense layers} followed by a linear output layer. Any two nodes of a dense layer act independently of each other and do not share weights and biases. Their input is given by the output of all nodes in the previous layer -- or all input features if it is the first layer -- and their output serves as an input to all nodes in the following layer, see Figure~\ref{fig:fc_nn} for a depiction of a small example.

FCNNs serve as powerful regression tools because they are able to represent large families of functions. In his 
\emph{Universal Approximation Theorem}, \citeA{Hor91} proves that FCNNs can approximate continuous functions on $\mathbb{R}$ arbitrarily well.

\begin{theorem}[Universal Approximation Theorem] 
	Let $N(\alpha)$ denote the space of functions that a fully connected neural network with activation function $\alpha: \mathbb{R} \to \mathbb{R}$, a single hidden layer with a finite number of neurons $l\in \mathbb{N}$ and a linear output layer can represent, i.e.
	\begin{align*}
		N(\alpha) = \left\{f: \mathbb{R}^d \to \mathbb{R} \mid f(x) = \sum_{i=1}^l w_i\alpha\left(\sum_{j=1}^d \bar{w}^{(i)}_j x_j + b^{(i)} \right) + b_i \right. \\ \left. \vphantom{\sum_{i}^{2}} \textrm{for some }w, b \in \mathbb{R}^l \textrm{ and }\bar{w}^{(i)}, \bar{b}^{(i)}  \in \mathbb{R}^d, 1 \leq i \leq l \right\}
	\end{align*}
	where $w, b \in \mathbb{R}^l$ are weights and biases of the output layer and $\bar{w}^{(i)}, \bar{b}^{(i)}  \in \mathbb{R}^d, 1 \leq i \leq l $ are the weights and biases of the $l$ individual neurons in the hidden layer.
	Assuming the activation function $\alpha: \mathbb{R} \to \mathbb{R}$ is non-constant, unbounded and continuous, $N(\alpha)$ is dense in $C(X)$ for compact $X\subseteq \mathbb{R}$ in the uniform topology, i.e. for any $f \in C(X)$ and arbitrary $\varepsilon > 0$, there is $g \in N(\alpha)$ such that 
	\begin{align*}
		\sup_{x \in X} |f(x) - g(x)| < \varepsilon.
	\end{align*}
\end{theorem}
 The \emph{Rectified Linear Unit (ReLU)} nonlinearity $\alpha: \mathbb{R} \to \mathbb{R}_+$ given by $\alpha(x) := \max(0, x)$ fulfills the conditions of being non-constant, unbounded and continuous and so in theory ReLU FCNNs allow for approximation of continuous functions to arbitrary accuracy. However, the reason the ReLU has become a de facto standard in recent years \cite{LBH15} is that in comparison to first generation nonlinearities such as the \emph{sigmoid} or \emph{tanh}, ReLU networks are superior in terms of their \emph{algorithmic learnability}, see more in Section \ref{sec:nn_training}.

\begin{remark}
	Over the years, various alternative activation functions have been proposed such as Leaky ReLU \cite{HZRS15}, ELU \cite{CUH15} or lately the SiLU \cite{EUD18, RZL17}. To date, none of these activation functions have been shown to consistently outperform ReLUs \cite{RZL17}, so a systematic comparison of the effect of different activation functions on training results has been left for future research.
\end{remark}

\section{Calibration of option pricing models}
\label{sec:calibration_general}
The implied volatility map $\varphi: \mathcal{I} \to \mathbb{R}_+$ defined in \eqref{eq:iv_map} formalizes the influence of model parameters on an option pricing model's implied volatility surface. \emph{Calibration} describes the procedure of tweaking model parameters to fit a model surface to an empirical IV surface obtained by transforming liquid European option market prices to Black-Scholes IVs (cf. Figure \ref{fig:spxiv150218}). A mathematically convenient approach consists of minimizing the weighted squared differences between market and model IVs of $N \in \mathbb{N}$ plain vanilla European options.

\begin{proposition}[Calibration objective]
	\label{prop:calibration_objective}
	Consider a (rough) stochastic volatility model with model parameters $\mu \in \mathcal{M} \subseteq \mathbb{R}^{m}$ and embedded market information $\xi \in \mathcal{E} \subseteq \mathbb{R}^{k}$ (recall~Def.~\ref{def:general_pricing_map}). Suppose the \emph{market} IV quotes of $N$ European options with moneyness $M^{(i)}$ and time to maturity $T^{(i)}$ are given by
	\begin{align*}
	\mathbf{Q} := \left(Q\left(M^{(1)}, T^{(1)}\right), \ldots, Q\left(M^{(N)}, T^{(N)}\right)\right)^T \in \mathbb{R}^N
	\end{align*}
	and analogously the \emph{model} IV quotes of the same options under said pricing model are given by
	\begin{align*}
	\bm{\varphi}\left(\mu, \xi\right) := \left(\varphi\left(\mu, \xi, M^{(1)}, T^{(1)}\right), \ldots, \varphi\left(\mu, \xi, M^{(N)}, T^{(N)}\right)\right)^T \in \mathbb{R}^N.
	\end{align*}
	Given market quotes $\bm{Q}$ and market information $\xi$, we define the residual $\bm{R}(\mu):~\mathcal{M}~\to~\mathbb{R}^N$ between market and model IVs by
	\begin{align*}
		\bm{R}(\mu) := \bm{\varphi}(\mu, \xi) - \mathbf{Q}
	\end{align*}
	so that the calibration objective becomes
	\begin{align}
	\label{eq:calibration_objective}
	\mu^\star &= \argmin_{\mu \in \mathcal{M}}  \norm{\mathbf{W}^\frac{1}{2}\bm{R}(\mu)}_2^2 = \argmin_{\mu \in \mathcal{M}} \norm{\mathbf{W}^\frac{1}{2}\left[\bm{\varphi}(\mu, \xi) - \mathbf{Q}\right]}_2^2  := \Psi\left(\mathbf{W},\xi, \bm{Q}\right)
	\end{align}
	where $\bm{W} = \textrm{diag}\left[w_1, \ldots, w_N\right] \in \mathbb{R}^{N \times N}$ is a diagonal matrix of weights and $\norm{\cdot}_2$ denotes the standard Euclidean norm.
\end{proposition}

Since $\bm{R}(\mu): \mathcal{M} \to \mathbb{R}^N$ is non-linear in the parameters $\mu  \in \mathcal{M} \subseteq \mathbb{R}^{m}$ and $N > m$, the optimization objective \eqref{eq:calibration_objective} is an example of an overdetermined non-linear least squares problem, usually solved numerically using iterative solvers such as the de-facto standard Levenberg-Marquardt (LM) algorithm \cite{Lev44, Mar63}. 

\begin{proposition}[LM calibration]
	Suppose $\bm{R}: O \to \mathbb{R}^N$ is twice continuously differentiable on an open set $O \subseteq \mathbb{R}^m$ and $N>m$. Let $\bm{J}:~O~\to~\mathbb{R}^{N \times m}$ denote the Jacobian of $\bm{R}$ with respect to the model parameters $\mu \in \mathbb{R}^m$, i.e. its components are given by
	\begin{align*}
	\left[\bm{J}_{ij}\right]_{\substack{1\le i\le N,\\[2pt] 1\le j\le m}} = \left[\frac{\partial \bm{R}_i(\mu) }{\partial \mu_j }\right]_{\substack{1\le i\le N,\\[2pt] 1\le j\le m}} = \left[ \frac{\partial \bm{\varphi}_i\left(\mu, \xi\right)}{\partial \mu_j}\right]_{\substack{1\le i\le N,\\[2pt] 1\le j\le m}}.
	\end{align*}
	With regards to the objective in \eqref{eq:calibration_objective}, the algorithm starts with an initial parameter guess $\mu_0 \in \mathbb{R}^m$ and then at each iteration step with current parameter estimate $\mu_k \in \mathbb{R}^m, k \in \mathbb{N}$, the parameter update $\Delta_\mu \in \mathbb{R}^m$ solves
	\begin{align}
	\label{eq:lev_mar}
	\left[\bm{J}(\mu_k)^T \bm{W}\bm{J}(\mu_k) + \lambda I_m\right] \Delta_\mu =  \bm{J}(\mu_k)^T \bm{W}\bm{R}(\mu_k)
	\end{align}
	where $I_m \in \mathbb{R}^{m\times m}$ denotes the identity and $\lambda \in \mathbb{R}$.
\end{proposition}

It is hence necessary that the \emph{normal equations} \eqref{eq:lev_mar} be quickly and accurately solved for the iterative step $\Delta_\mu$. In a general (rough) stochastic volatility setting this is problematic: The true implied volatility map $\varphi:  \mathcal{I} \to \mathbb{R}_+$ as well as its Jacobian $\bm{J}:~O~\to~\mathbb{R}^{N \times m}$ are unknown in analytical form. In the absence of an analytical expression for $\Delta_\mu$, an immediate remedy is:
\begin{itemize}
	\item[(I)] Replace the (theoretical) true pricing map $P_0: \mathcal{I} \to \mathbb{R}_+$ defined in \eqref{eq: pricing_map_p} by an efficient numerical approximation $\tilde{P}_0: \mathcal{I} \to \mathbb{R}_+$ such as Monte Carlo, Fourier Pricing or similar means. This gives rise to an approximate implied volatility map $\tilde{\varphi}: \mathcal{I} \to \mathbb{R}_+$.
	\item[(II)] Apply finite-difference methods to $\tilde{\varphi}: \mathcal{I} \to \mathbb{R}_+$ to compute an approximate Jacobian $\tilde{\bm{J}}:~O~\to~\mathbb{R}^{N \times m}$.
\end{itemize}
In many (rough) stochastic volatility models such as \emph{rough Bergomi}, expensive Monte Carlo simulations have to be used to approximate the pricing map. In a common calibration scenario where the normal equations \eqref{eq:lev_mar} have to be solved frequently, the approach outlined above thus renders calibration prohibitively expensive.

\subsection{Deep calibration}
\label{subsec:deep_cal}
\begin{algorithm}[]
	\label{algo:levmar_nn}
	\SetAlgoLined
	\KwIn{Implied vol map $\bm{\varphinn}$ and its Jacobian $\bm{J}_\textrm{NN}$, market quotes $\bm{Q}$, market info $\xi$}
	\SetKwInOut{Parameters}{Parameters}
	\Parameters           {Lagrange multiplier $\lambda_0 > 0$, maximum number of iterations $n_\textrm{max}$, minimum tolerance of step norm $\varepsilon_\textrm{min}$, bounds $0 < \beta_0 < \beta_1 < 1$}
	\KwResult{Calibrated model parameters $\mu^\star$}
	initialize model parameters $\mu = \mu_0$ and step counter $n=0$\;
	compute $\bm{R}(\mu) = \bm{\varphinn}(\mu, \xi) - \mathbf{Q}$ and $\bm{J}_\textrm{NN}(\mu)$ and solve normal equations \eqref{eq:lev_mar} for $\Delta_\mu$\;
	\While{$n < n_\textrm{max}$ and $\norm{\Delta_\mu}_2 > \varepsilon$}{
		compute relative improvement $
		c_\mu = \frac{\norm{\bm{R}(\mu)}_2 - \norm{\bm{R}(\mu + \Delta_\mu)}_2}{\norm{\bm{R}(\mu)}_2 - \norm{\bm{R}(\mu) + \bm{J}_\textrm{NN}(\mu) \Delta_\mu}_2}$ with respect to predicted improvement under linear model\;
		\lIf{$c_\mu \leq \beta_0$}{
			reject $\Delta_\mu$, set $\lambda = 2\lambda$}
		\lIf{$c_\mu \geq \beta_1$}{
			accept $\Delta_\mu$, set $\mu = \mu + \Delta_\mu$ and $\lambda = \frac{1}{2}\lambda$}
		compute $\bm{R}(\mu)$ and $\bm{J}_\textrm{NN}(\mu)$ and solve normal equations \eqref{eq:lev_mar} for $\Delta_\mu$\;
		
		set $n = n + 1$\;
	}
	\caption{Deep calibration (LM combined with NN regression)}
\end{algorithm}

In a first step, we use the approximate implied volatility map $\tilde{\varphi}: \mathcal{I} \to \mathbb{R}_+$ to synthetically generate a large and as accurate as computationally feasible set of labeled data 
\begin{align*}
\mathcal{D} := \left\{\left(x^{(i)}, \tilde{\varphi}\left(x^{(i)}\right)\right) \mid x \in \mathcal{I}\right\}_{i=1}^n \in (\mathcal{I}\times \mathbb{R}_+)^n,\quad  n \in \mathbb{N}.
\end{align*}
Here, it is sensible to trade computational savings for an increased numerical accuracy since the expensive data generation only has to be performed once. Using the sample input-output pairs $\mathcal{D}$, a ReLU FCNN is trained to approximate $\tilde{\varphi}: \mathcal{I} \to \mathbb{R}_+$, in other words, we use a ReLU FCNN to regress response variables $\tilde{\varphi}\left(x^{(i)}\right) = \tilde{\varphi}\left(\mu^{(i)}, \xi^{(i)}, M^{(i)}, T^{(i)}\right)$ on explanatory variables $\left(\mu^{(i)}, \xi^{(i)}, M^{(i)}, T^{(i)}\right)$.  We denote this function that the network is now able to represent by $\varphi_\textrm{NN}: \mathcal{I} \to \mathbb{R}_+$. With respect to the repeated solving of the normal equations \eqref{eq:lev_mar}, the benefit of this new approach is twofold:
\begin{itemize}
	\item[(I)] Evaluations of $\varphi_\textrm{NN}: \mathcal{I} \to \mathbb{R}_+$ amount to forward runs of a trained ReLU FCNN. Computationally, forwards runs come down to highly optimized and parallelizable matrix-matrix multiplications combined with element-wise comparison operations -- recall the ReLU activation is given by $\alpha(\cdot) = \max(0,\cdot)$ -- both of which are fast.
	\item[(II)] In order to perform \emph{backpropagation}, the standard training algorithm for neural networks, industrial grade machine learning software libraries such as Google Inc.'s Tensorflow \cite{ABC16} ship with built-in implementations of \emph{automatic differentiation} \cite{BPBR15}. This may easily be exploited to quickly compute approximative Jacobians  $\bm{J}_\textrm{NN}:~O~\to~\mathbb{R}^{N \times m}$ accurate to machine precision.
\end{itemize}
It is also important to stress that trained networks can be efficiently stored, moved and loaded, so training results can be shared and deployed quickly.
\begin{remark}
	\citeA{Her17} calibrates the \citeA{HW90} short-rate model by directly learning calibrated model parameters from market data, i.e. the total calibration routine $\Psi$ in \eqref{eq:calibration_objective}. Extending his approach to equity models necessitates a network topology that allows to learn from empirical IV point clouds. Here, adaptations of Convolutional Neural Networks (CNNs) invented for computer vision problems might be worthwhile to explore.
\end{remark}

\section{Neural network training}
\label{sec:nn_training}
While theoretically easy to understand, the training of neural networks in practice often becomes a costly and most importantly time--consuming exercise full of potential pitfalls.
To this end, we outline here the approach taken in this paper, briefly mentioning important \emph{tricks of the trade} that have been utilized to facilitate or accelerate the training of the ReLU FCNN networks. 
\subsection{Generation of synthetic labeled data}
\label{sec:generation_synthetic_labdata}
\begin{table}
	\caption{Marginal priors of model parameters $\mu$ for synthetically generating $\mathcal{D}$. The continuous uniform distribution on the interval bounded by $a_i, b_i \in \mathbb{R}$ is denoted by $\mathcal{U}[a_i, b_i]$ and $\mathcal{N}_\textrm{trunc}[a_i,b_i, \lambda , \sigma]$ stands for the normal distribution with mean $\lambda \in \mathbb{R}$ and standard deviation $\sigma \in \mathbb{R}_+$, truncated to the interval $[a_i, b_i]$ with $a_i, b_i \in \mathbb{R}$.}
	\label{ta:sampling_ranges}
	\centering
	\begin{tabular}{cccc}
		\toprule
		\multicolumn{2}{c}{Heston} &
		\multicolumn{2}{c}{rough Bergomi} \\
		\cmidrule{1-4}
		Parameter & Marginal & Parameter & Marginal \\ 
		\midrule 
		$\eta$ & $\mathcal{U}[0, 5]$ & $\eta$ & $\mathcal{N}_\textrm{trunc}[1,4, 2.5, 0.5]$\\
		$\rho$ & $\mathcal{U}[-1, 0]$ & $\rho$ & $\mathcal{N}_\textrm{trunc}[-1,-0.5, -0.95, 0.2]$\\
		$\lambda$ & $\mathcal{U}[0, 10]$ & $H$ &  $\mathcal{N}_\textrm{trunc}[0.01,0.5, 0.07, 0.05]$\\
		$\bar{v}$ & $\mathcal{U}[0, 1]$ & $v_0$ & $\mathcal{N}_\textrm{trunc}[0.05,1, 0.3, 0.1]^2$\\
		$v_0$ & $\mathcal{U}[0, 1]$ & \\ 
		\bottomrule
	\end{tabular}
\end{table}

The ability of a neural network to learn the implied volatility map $\varphi$ to a high degree of accuracy critically hinges upon the provision of a large and accurate labeled data set
\begin{align*}
	\mathcal{D}= \left\{\left(\mu^{(i)}, \xi^{(i)}, M^{(i)}, T^{(i)}, \tilde{\varphi}\left(\mu^{(i)}, \xi^{(i)}, M^{(i)}, T^{(i)}\right)\right)\right\}_{i=1}^n \in(\mathcal{I}\times \mathbb{R}_+)^n, \quad n \in \mathbb{N}.
\end{align*}
Knowledge of the parametric dependence structure $\tilde{\varphi}: \mathcal{I} \to \mathbb{R}_+$ between inputs and corresponding outputs allows us to address these requirements adequately. First, trading computational savings for increased numerical accuracy, we ensure that $\norm{\tilde{\varphi} - \varphi}_\infty < \epsilon$ for $\epsilon$ small.
Second, we can sample an arbitrarily large set of labeled data $\mathcal{D}$, allowing the network to learn the underlying dependence structure $\tilde{\varphi}$ -- rather than noise present in the training set -- and generalize well to unseen test data. In the numerical tests in Section \ref{sec:numerical_results}, we draw $n=|\mathcal{D}|= 10^6$ iid sample inputs from a to be specified sampling distribution $\mathcal{G}$ on $\mathcal{I}$ and compute the corresponding outputs as follows: For Heston, we use the Fourier pricing method implemented in the open-source quantitative finance library \emph{QuantLib} \cite{AB15} which makes use of the well-known fact that the characteristic function of the log asset price is known. For \emph{rough Bergomi}, we use a self-coded, parallelized implementation of a slightly improved version of the Monte Carlo scheme proposed by \citeA{MP18}. Black-Scholes IVs are inverted from option prices using a publicly available implementation of the implied volatility solver by \citeA{Jac15}. The full dataset $\mathcal{D}$ is then randomly shuffled and partitioned into training, validation and test sets $\mathcal{D}_{train},\mathcal{D}_{valid}$ and $ \mathcal{D}_{test}$ of sizes $n_\textrm{train}, n_\textrm{valid}$ and $n_\textrm{test}$ respectively.\footnote{ The code has been made available at \href{https://github.com/roughstochvol}{https://github.com/roughstochvol}.} 

An important advantage of being able to synthetically generate labeled data is the freedom in choosing the sampling distribution $\mathcal{G}$ on $\mathcal{I}$. Prior to calibration, little is known about the interplay of model parameters and particular model parameter regions of highest interest to be learned accurately. Consequently, we assume zero prior knowledge of the (joint) relevance of model parameters $\mu$ in the Heston experiment in Section \ref{sec:numerical_results}. An ad-hoc approach is to sample individual model parameters independently of each other from the uniformly continuous marginal distributions collected in Table \ref*{ta:sampling_ranges}. A similar reasoning also applies in the rough Bergomi experiment in Section \ref{sec:numerical_results}, except that here we do assume some prior marginal distributional knowledge and use truncated normal marginals instead of uniform marginals.

On the other hand, it is reasonable to increase the number of samples in option parameter regions with high liquidity since these are given more weight by the calibration objective \eqref{eq:calibration_objective} and as such require to be more accurate.
To that end, we postulate a joint distribution of moneyness and time to maturity based on liquidity and estimate it using a weighted Gaussian kernel density estimation (wKDE) \cite{Sco15}: Let $L_i$ denote the market liquidity of an option $i, i \in \mathbb{N},$ with time to maturity $T^{(i)}$ and moneyness $M^{(i)}$. We proxy liquidities by inverse bid-ask spreads of traded European Call Options on SPX and then run a wKDE on samples $\left\{\left(M^{(i)}, T^{(i)}\right)\right\}_{i=1}^n$ with weights $\{L_i\}_{i=1}^n$ and a smoothing bandwidth. In a similar vein, one may also derive a multivariate distribution $\mathcal{K}_{\xi}$ of external market data $\xi \in \mathbb{R}^k$. 

With regards to the individual marginals collected in Table \ref{ta:sampling_ranges},
the sampling distribution $\mathcal{G}_\textrm{Heston}$ on $\mathcal{I} \subseteq \mathbb{R}^{m+k+2}$ is given by
\begin{align}
\label{eq:def_prob_I}
\mathcal{G}_\textrm{Heston} := \mathcal{U}^{\otimes m}[a_i, b_i] \otimes \mathcal{K}_{\xi} \otimes \mathcal{K}_{(M,T)} 
\end{align}
and analogously for the rough Bergomi model, we have
\begin{align}
\label{eq:rb_prior}
\mathcal{G}_\textrm{rBergomi} := \mathcal{N}_\textrm{trunc}^{\otimes m}[a_i, b_i, \lambda_i, \sigma_i] \otimes \mathcal{K}_{\xi} \otimes \mathcal{K}_{(M,T)}.
\end{align}

\subsection{Backpropagation and hyper parameter optimization}
Consider a ReLU FCNN with $L\in \mathbb{N}$ hidden layers as described in Section \ref{subsec:fcnn}. Let $n_l, 1\leq l \leq L$ denote the number of nodes of the hidden layers and $\mathcal{S}_{h_{\textrm{model}}}$ the function space spanned by such a network with model hyper parameters $h_\textrm{model} = (L, n_1, \ldots, n_{L})$. Let $X: \Omega \to \mathcal{I}$ denote a random input and consider $h_\textrm{model}$ fixed. Then the fundamental objective of neural network training is to learn a function that minimizes the generalization error:
\begin{align}
	f^\star_{h_{\textrm{model}}} = \argmin_{f_{h_{\textrm{model}}} \in \mathcal{S}_{h_{\textrm{model}}}} \norm{f_{h_{\textrm{model}}}(X) - \tilde{\varphi}(X)}^2_{L^2(\Omega)}, \quad X \sim \mathcal{G}\label{eq:train_obj_gen}
\end{align}
where $\mathcal{G} \in \{\mathcal{G}_\textrm{Heston}, \mathcal{G}_\textrm{rBergomi}\}$, depending on experiment.
In many calibration scenarios, $\tilde{\varphi}$ is a Monte-Carlo approximation to $\varphi$, so $\tilde{\varphi}(\cdot) = \varphi(\cdot) + \varepsilon$ for $\varepsilon$ some homoskedastic error with $\mathbb{E}(\varepsilon) = 0$ and $\textrm{Var}(\varepsilon) = \sigma^2>0$. The MSE loss in \eqref{eq:train_obj_gen} admits the well-known bias-variance decomposition 
\begin{align}
\label{eq:bias_var_decomp}
	\norm{f_{h_{\textrm{model}}}(X) - \tilde{\varphi}(X)}^2_{L^2(\Omega)} = \left(\mathbb{E}\left[f_{h_{\textrm{model}}}(X) - \varphi(X)\right]\right)^2 + \textrm{Var}\left[f_{h_{\textrm{model}}}(X)\right] + \sigma^2 
\end{align}
where in addition to a bias and variance term we also have the variance of the sample error, the irreducible error. The empirical analogue to \eqref{eq:train_obj_gen} relevant for practical training is given by
\begin{align}
\label{eq:train_obj_emp}
	f^\star_{h_{\textrm{model}}} \approx \argmin_{f_{h_{\textrm{model}}} \in \mathcal{S}_{h_{\textrm{model}}}} \frac{1}{n_\textrm{valid}} \sum_{i=1}^{n_\textrm{valid}} \left[f_{h_{\textrm{model}}}\left(x^{(i)}\right) - \tilde{\varphi}\left(x^{(i)}\right)\right]^2 
\end{align}
where $\left(x^{(i)}, \tilde{\varphi}\left(x^{(i)}\right)\right) \in \mathcal{D}_\textrm{valid}$. 
The optimization in the function space $\mathcal{S}_{h_{\textrm{model}}}$ corresponds to a high-dimensional nonlinear optimization in the space of network weights and biases, similarly to \eqref{eq:calibration_objective} typically addressed by gradient-based schemes. \emph{Backpropagation} \cite{GBC16}, a specific form of \emph{reverse-mode automatic differentiation} \cite{BPBR15} in the context of neural networks, prevails as the go-to approach to iteratively compute gradients of the empirical MSE loss with respect to weights and biases of all nodes in the network. The gradients are then often used in the well-known \emph{Mini-Batch Gradient Descent} \cite{GBC16} optimization algorithm, a variant of which called \emph{Adam} \cite{KB14} we use in our experiments. \emph{Adam} incorporates momentum to prevent the well-known zigzagging of \emph{Gradient Descent} in long and sharp valleys of the error surface and adaptively modifies a given global step size for each component of the gradient individually to speed up the optimization process. It in turn has its own optimization hyper parameters $h_\textrm{opt} = (\delta, \beta)$ where $\delta$ denotes the mentioned global learning rate and $\beta$ denotes the mini-batch size used. In the following, we denote the learning algorithm \emph{Adam} mapping training data $\mathcal{D}_\textrm{train}$ to a local minimizer $f^\star_{h_{\textrm{model}}}$ of \eqref{eq:train_obj_emp} by $\mathcal{A}_{h_\textrm{opt}}: \mathcal{I}^n \to \mathcal{S}_{h_{\textrm{model}}}$. 

Up to know, we treated the hyper parameters $(h_\textrm{opt}, h_\textrm{model})$ as fixed whereas in reality they may be varied and have a crucial influence on the training outcome. Indeed, for all other variables besides $(h_\textrm{opt}, h_\textrm{model})$ fixed, let us define a hyper parameter response function $\mathcal{H}$ by
\begin{align}
\mathcal{H}(h_\textrm{opt}, h_\textrm{model}) := \frac{1}{n_\textrm{valid}} \sum_{i=1}^{n_\textrm{valid}} \left[\left[\mathcal{A}_{h_\textrm{opt}}\left(\mathcal{D_\textrm{train}}\right)\right]_{h_{\textrm{model}}}\left(x^{(i)}\right) - \tilde{\varphi}\left(x^{(i)}\right)\right]^2.
\end{align}
In practice, it then turns out the real challenge in training neural networks to high accuracy lies in the additional (outer) optimization over hyper parameters:
\begin{align}
\label{eq:hyperparam_response_function}
	(h_\textrm{opt}^\star, h_\textrm{model}^\star)
	= \argmin_{\left(h_\textrm{opt}, h_\textrm{model}\right)} \mathcal{H}(h_\textrm{opt}, h_\textrm{model}).
\end{align}
The scope of effect of hyper parameters $h_\textrm{model}$ and $h_\textrm{opt}$ does not overlap: The former determines the capacity of $\mathcal{S}_{h_{\textrm{model}}}$, the latter governs which local minimizer  $f^\star_{h_{\textrm{model}}}$ the optimization algorithm $\mathcal{A}_{h_\textrm{opt}}$ converges to and the speed with which this happens. This allows us to treat their optimization separately. A coarse grid search reveals that adding additional layers beyond 4 hidden layers does not consistently reduce errors on the validation set. Rather, networks become harder to train as evidenced by errors fluctuating more wildly on the validation set. We suspect this is a consequence of what \citeA{IS15} call \emph{internal covariate shift}: First-order methods such as Gradient Descent are blind to changes in the weights and biases of the layers feeding into a given layer and so with deeper networks the propagating and magnifying effects of changes in one layer to subsequent layers worsen and slow down the training. On the other hand, our locally available compute resources max out at $4096=2^{12}$ nodes per layers, so we fix $h_\textrm{model} = (4) \times (4096)^4$.
\begin{figure*}
	\label{fig:nn_topology}
	\begin{center}
		\includegraphics[width=.7\textwidth]{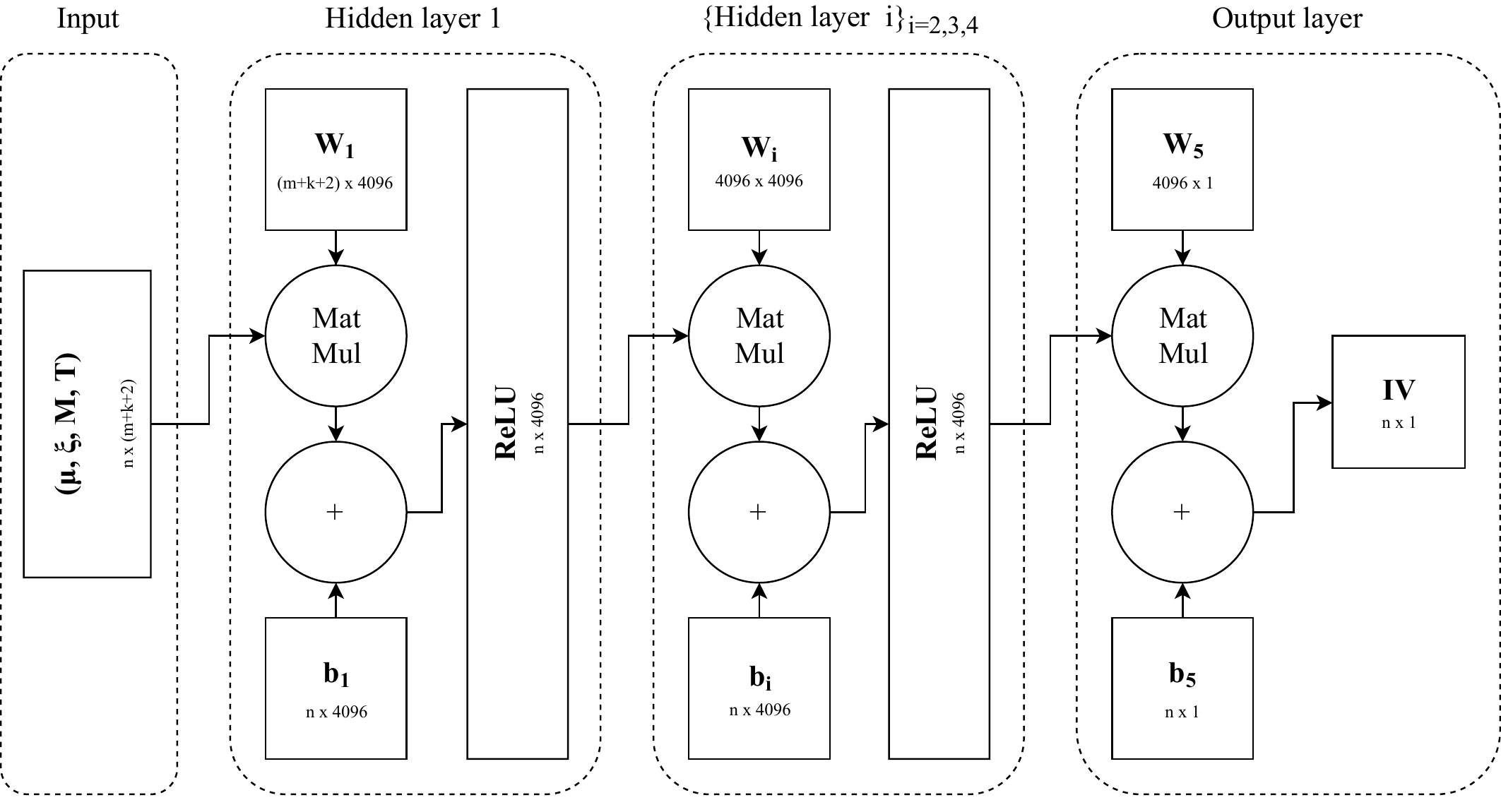}
		\caption{\textbf{Schematic of IV ReLU FCNN}. Depiction of 4-layer ReLU FCNN used to learn IV maps. It consists of $2^{12} = 4096$ nodes at each hidden layer. Note the output layer is a linear layer with no activation function. Rectangles denote tensors and circles denote operations, \emph{MatMul} is matrix multiplication. The number of parallel IV calculations is given by $n\in \mathbb{N}$.}
	\end{center}
\end{figure*}
Each evaluation of the hyper parameter response function $\mathcal{H}$ in \eqref{eq:hyperparam_response_function} requires a ReLU FCNN to be fully trained from scratch which is a very costly operation in terms of time and computing resources. Moreover, gradient-based optimization approaches are ruled out by the fact that gradients of $\mathcal{H}$ with respect to $h_\textrm{opt}$ are unavailable (after all, batch sizes are discrete). In our experiments, we explore the use of Gaussian Regression \cite{RW06, SLA12} which is an adaptive gradient-free minimization algorithm. Postulating a surrogate Gaussian model for $\mathcal{H}$, it takes existing function evaluations into account and -- balancing exploitation and exploration -- iteratively proposes the next most promising candidate input in terms of information gain. As is common in applied sciences, we use a Mat\'{e}rn Kernel for the covariance function of the Gaussian model and the Lower Confidence Bound (LCB) acquisition function.

\subsubsection{Tricks of the trade}

\textbf{Feature scaling} or \textbf{preconditioning} is a standard preprocessing technique applied to input data of optimization algorithms in order to improve the speed of optimization. After the data set $\mathcal{D}$ has been partitioned into training, validation and test sets, we compute the sample mean $\bar{x}_\textrm{train} \in \mathbb{R}^{m+k+2}$ of the inputs across the training set  and the corresponding sample standard deviation $s_\textrm{train} \in \mathbb{R}^{m+k+2}$. For each input $x^{(i)} \in \mathcal{I}$ from $\mathcal{D}$, its standardized version $\hat{x}^{(i)}$ is given by
\begin{align*}
	\hat{x}^{(i)} := \frac{x^{(i)} - \bar{x}_\textrm{train}}{s_\textrm{train}}, \quad 1\leq i \leq n.
\end{align*}
where the operations are defined componentwise.
We then use these standardized inputs $\hat{x}^{(i)}$ -- which have zero offset and unit scale -- for training and prediction. It is important to stress that \emph{all} $n$ inputs from the complete set $\mathcal{D}$ are standardized using the \emph{training} mean and standard deviation, including those of the validation and test sets.

\textbf{Weight initialization} is an important precursor to the iterative optimization process of \emph{Adam}. Initialization is a delicate task that may speed up or hamper the training process all together: If within (but not necessarily across) all layers, weights and biases of all nodes are identical, then the same is true for their outputs and the partial derivative of the loss with respect to their weights and biases, impeding any learning on the part of the optimizer. To \emph{break the symmetry}, it is standard procedure to draw weights from a symmetric probability distribution centered at zero. 
Suppose $w_{ij}^{(l)}$ denotes the weight of node $i, 1 \leq i \leq  n_l$ in layer $1 \leq l \leq L$ being multiplied with the output of node $j, 1 \leq j \leq n_{l-1}$ in layer $l-1$ and $n_0$ denotes the number of network inputs. \citeA{HZRS15} suggest the weights and biases be independently drawn as follows
\begin{align*}
w_{ij}^{(l)} \sim \mathcal{N}\left(0, \frac{2}{n_{l-1}}\right), \quad 
b_i^{(l)} = 0.
\end{align*}
Adapting an argument by \citeA{GB10} for linear layers to ReLU networks, they can show that -- under some assumptions -- this ensures that, at least at initialization, input signals and gradients do not get magnified exponentially during forward or backward passes of backpropagation.

\textbf{Regularization}  in the context of regression -- be it deterministic in the case of $L^2$ or $L^1$ or stochastic in the form of Dropout \cite{SHK14} -- describes a set of techniques aimed at modifying a training algorithm so as to reduce overfitting on the training set. With regards to \eqref{eq:bias_var_decomp}, the conceptual idea is that a modified optimizer allows to trade an increased bias of the estimator for an over proportional decrease in its variance, effectively reducing the MSE overall. In our experiments, we only regularize in time in the form of \emph{early stopping}: While optimizing the weights and biases on the training set, we periodically check the performance on the validation set and save the model if a new minimum error is reached. When the error on the validation set begins to stall, training is stopped.

\textbf{Batch normalization} (BN) devised by \citeA{IS15} is very popular technique to facilitate and accelerate the training of deeper networks by addressing the mentioned \emph{internal covariate shift}. It alters a network's topology by inserting normalization operations between linearities and non-linearities of each dense layer, effectively reducing the dependence of each node's input on the weights and biases of all nodes in previous layers. Our numerical experiments confirm a strongly regularizing effect of BN as is well-known in the literature, reducing the expressiveness of our networks considerably and hence leading to worse performance. Despite its success in allowing to train deeper networks, we hence decided to turn it off.

\section{Numerical experiments}
\label{sec:numerical_results}

\begin{table}
	\caption{Reference model parameters $\mu^\dagger$ for Heston and rough Bergomi. Obtained from  (Gatheral, 2011) and (Bayer et al., 2016) respectively.}
	\label{ta:ref_parameter}
	\centering
	\begin{tabular}{lrlr}
		\toprule
		\multicolumn{2}{c}{Heston} &
		\multicolumn{2}{c}{rough Bergomi} \\
		\cmidrule{1-4}
		Parameter & Value & Parameter & Value \\ 
		\midrule 
		
		$\eta$ & 0.3877 &  $\eta$ & 1.9\\
		$\rho$ & -0.7165 & $\rho$ & -0.9 \\
		$\lambda$ & 1.3253 &  $H$ & 0.07\\ 
		$\bar{v}$ & 0.0354 & $v_0$ & 0.01 \\
		$v_0$ & 0.0174 & \\
		\bottomrule
	\end{tabular}
	
\end{table}

Here, we examine the performance of our approach by applying it to the option pricing models recalled in Section \ref{sec:background}: First, we consider the Heston model as a test case and then the rough Bergomi model as a representative from the class of \emph{rough} stochastic volatility models. Specifically, we look at the speed and accuracy of the learned implied volatility map $\varphi_\textrm{NN}: \mathcal{I} \to \mathbb{R}_+$. A systematic comparison of performance metrics between existing methods and our approach has been left for future research.

The Gaussian hyper parameter optimization and individual network training runs are performed on a local CPU-only compute server. Unless otherwise stated, all computations and performance measures referenced in this section are performed on a standard early 2015 Apple Mac Book with a 2.9 GHz Intel Core i5 CPU with no GPU used.

\subsection{The Heston model}

\begin{sidewaysfigure}[p] 
	\begin{subfigure}[b]{0.5\linewidth}
		\centering
		\includegraphics[width=\linewidth]{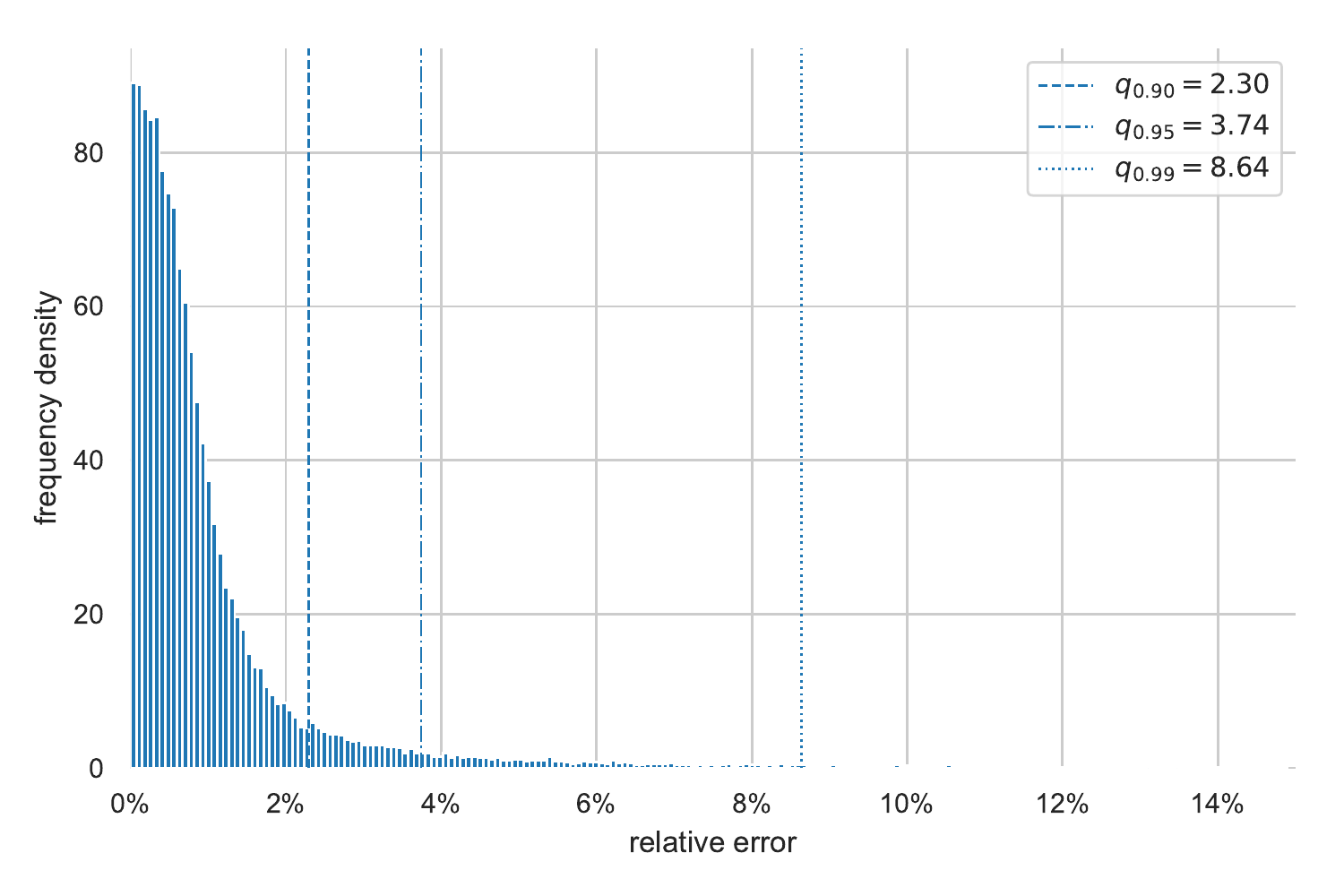} 
		\caption{Normalized histogram of RE \eqref{def:rel_err_iv} on testset with quantiles.} 
		\label{fig:heston_comp_graphs_c} 
	\end{subfigure}
	\begin{subfigure}[b]{0.5\linewidth}
		\centering
		\includegraphics[width=\linewidth]{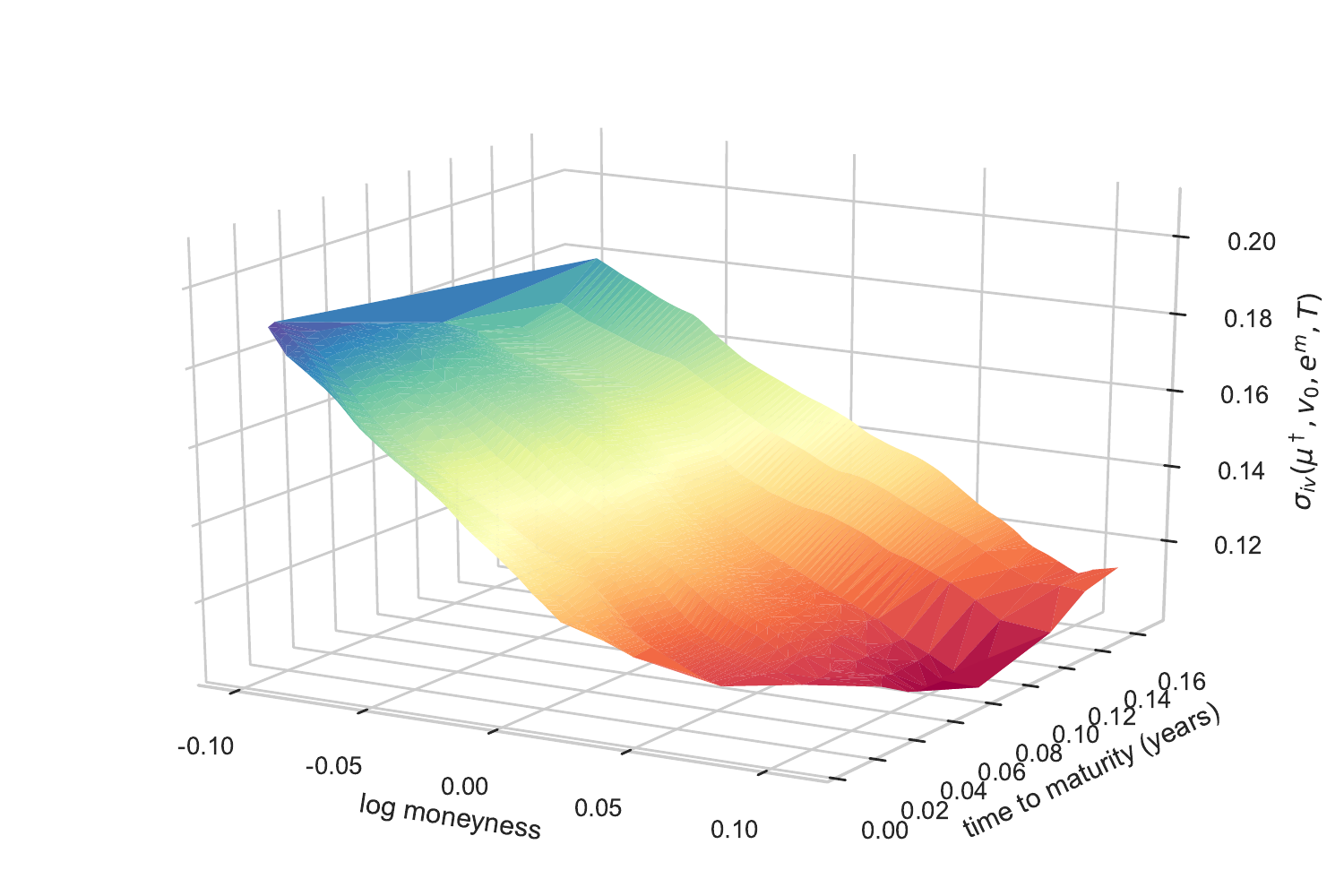} 
		\caption{Heston IV surface computed using learned IV map $\varphi_\textrm{NN}$.} 
		\label{fig:heston_comp_graphs_a} 
	\end{subfigure}
	\begin{subfigure}[b]{0.5\linewidth}
		\centering
		\includegraphics[width=\linewidth]{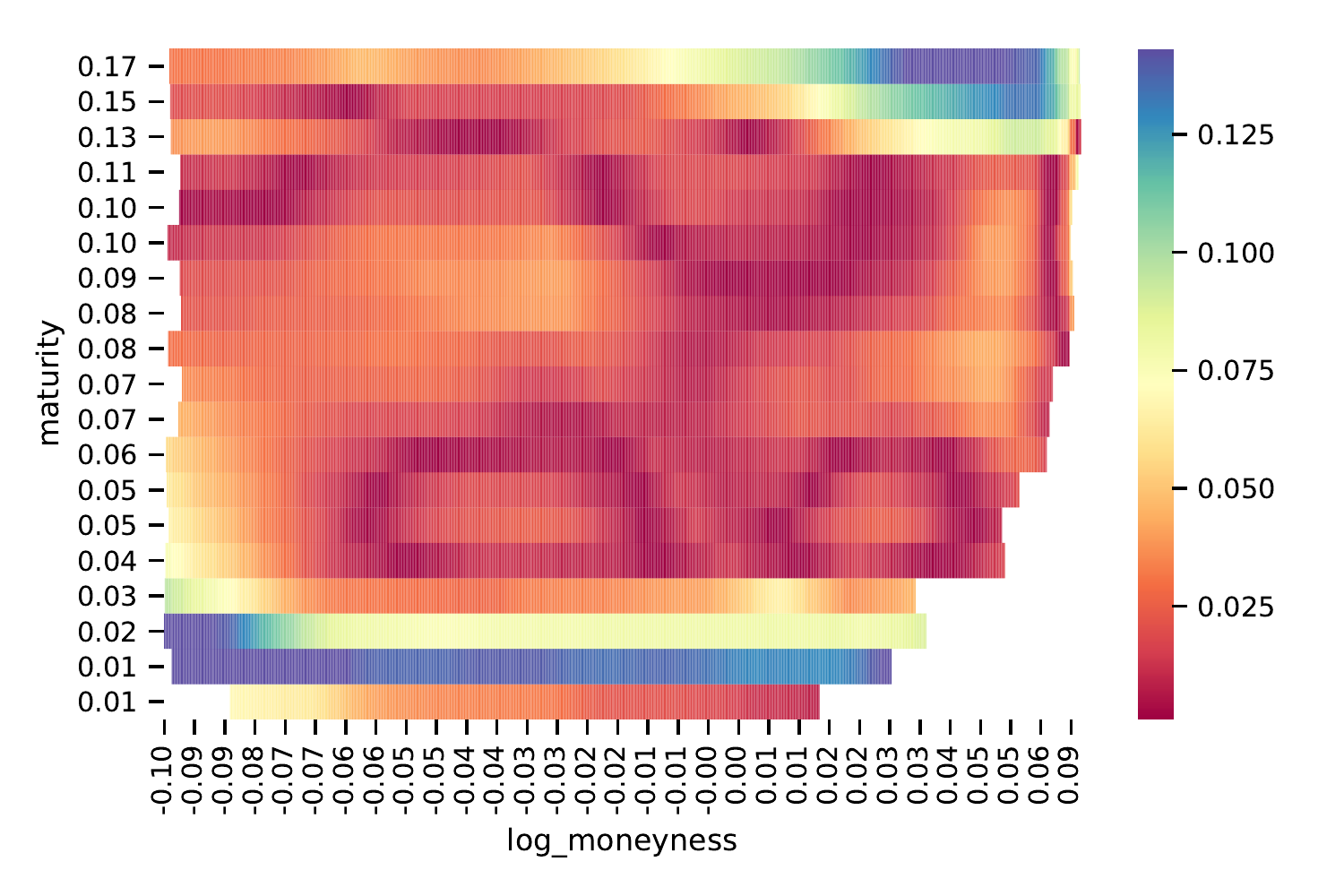} 
		\caption{Interpolated heatmap of $\textrm{RE}(\mu^\dagger, m, T)$ \eqref{def:rel_err_iv} for varying $m,T$.} 
		\label{fig:heston_comp_graphs_b} 
	\end{subfigure}
	\begin{subfigure}[b]{0.5\linewidth}
		\centering
		\includegraphics[width=\linewidth]{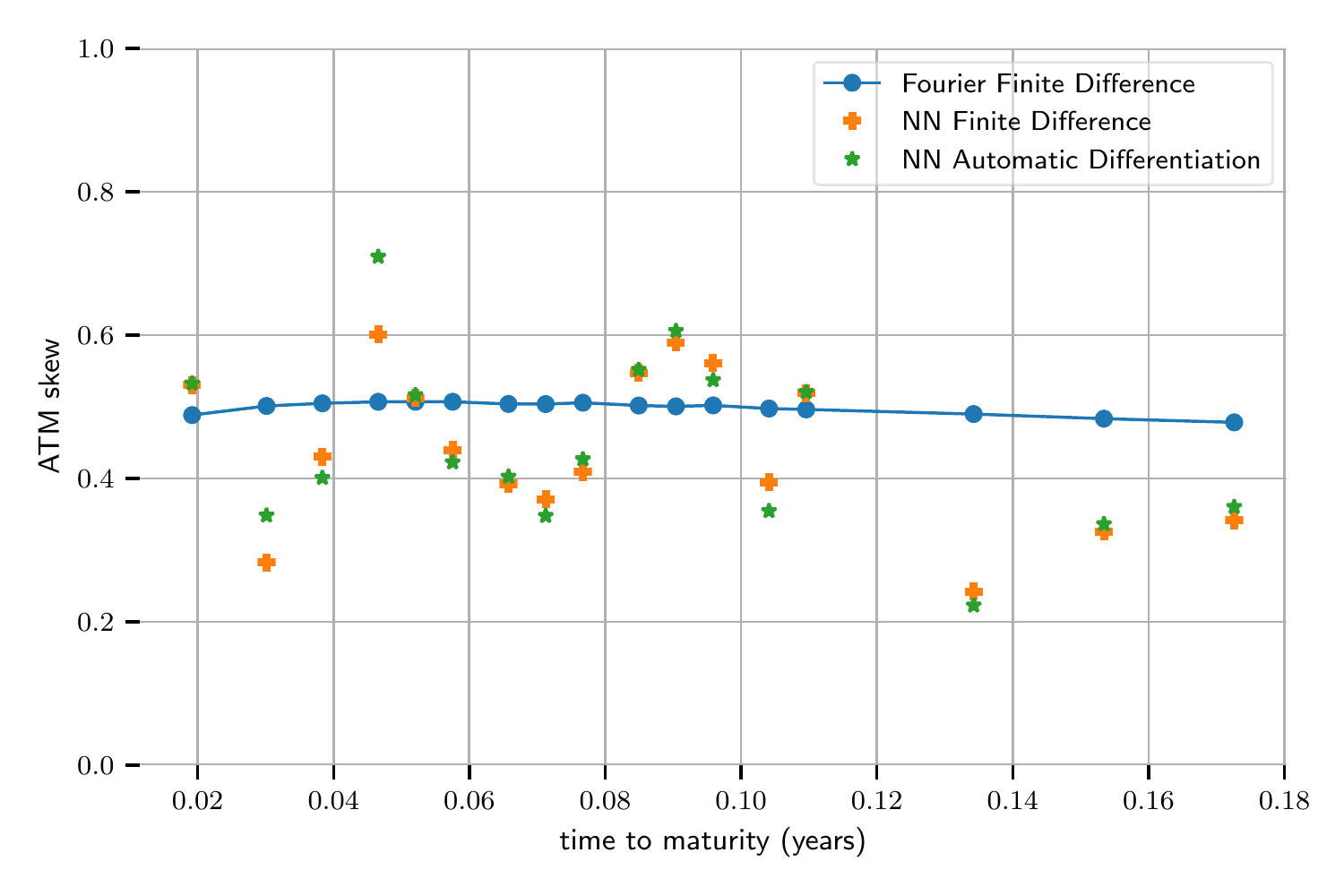} 
		\caption{Approximations to the ATM volatility skew.} 
		\label{fig:heston_comp_graphs_d} 
	\end{subfigure} 
	\caption{\textbf{Accuracy of the IV map $\varphinn$ as learned by the Heston ReLU FCNN.}}
	\label{fig7} 
\end{sidewaysfigure}
Following the approach outlined in Section \ref{sec:generation_synthetic_labdata}, we estimate $\mathcal{K}_{(M,T)}$ using SPX Option Price data\footnote{ Option prices for SPX Weeklys can be retrieved from a publicly available database at \href{www.cboe.com/DelayedQuote/QuoteTableDownload.aspx}{www.cboe.com/DelayedQuote/QuoteTableDownload.aspx}.} from 15th February 2018. Empirically, we observe that a majority of the liquidity as proxied by inverse bid-ask spreads is concentrated in the small region given by $- 0.1 \leq  m \leq 0.28$ and $\frac{1}{365} \leq T \leq 0.2$ which is why for this test case we exclusively learn the IV map on this bounded domain. The size of the labeled set data $\mathcal{D}$ is $n=990000$ of which we allocate $n_\textrm{train} = 900000$ samples to the training set and $n_\textrm{valid} = n_\textrm{test} = 45000$ to test and validation sets.

Single evaluations of the learned implied volatility map $\varphi_\textrm{NN}:\mathcal{I} \to \mathbb{R}_+$ and the associated Jacobian $\bm{J}_\textrm{NN}:~O~\to~\mathbb{R}^{N \times m}$ are extremely fast with about $36$ms on average to compute both together, making this neural network based approach at least competitive with existing Fourier-based schemes. To determine the accuracy of $\varphinn$, we define
\begin{align}
\label{def:rel_err_iv}
\textrm{RE}(\mu, m,T) := \frac{|\varphi_\textrm{NN}(\mu, v_0, e^m, T) - \tilde{\varphi}(\mu, v_0, e^m, T)|}{\tilde{\varphi}(\mu, v_0, e^m, T)}
\end{align}
to be the relative error of the output of $\varphinn$ with respect to that of a Fourier-based reference map $\varphitilde$ for model parameters $\mu$, option parameters $(m,T)$ and fixed spot variance $v_0$. Figure \ref{fig:heston_comp_graphs_c} shows a normalized histogram of relative errors on the test set
where $\mu$ and $(M,T)$ are allowed to vary across samples, demonstrating that empirically, $\varphinn$ approximates $\varphitilde$ with a high degree of accuracy. In typical pricing or calibration scenarios, we are interested in the accuracy of $\varphinn$ for some fixed model parameters $\mu$ which is why in Figures \ref{fig:heston_comp_graphs_a}, \ref{fig:heston_comp_graphs_b} and \ref{fig:heston_comp_graphs_d}, we fix $\mu = \mu^\dagger$ with $\mu ^\dagger$ the reference model parameters in Table \ref{ta:ref_parameter}. In Figure \ref{fig:heston_comp_graphs_a}, we compute an IV point cloud using $\varphi_\textrm{NN}$, interpolate it using a (not necessarily arbitrage-free) Delaunay triangulation and recover a characteristic Heston-like model IV surface. Indeed, as the heatmap of interpolated relative errors in Figure \ref{fig:heston_comp_graphs_b} shows, these are small across most of the IV surface, with increased relative errors only for times on the short and long end which may be attributed to less training because of less liquidity. Finally, in Figure \ref{fig:heston_comp_graphs_d}, we plot three different approximations to the Heston ATM volatility skew for small times: A reference skew in blue obtained by a finite difference approximation using $\varphitilde$, another skew in orange obtained by the same method but applied to $\varphinn$ and finally the exact ATM skew of $\varphinn$ in green, available by automatic differentiation. As is to be expected, $\varphinn$ recovers the characteristic flat behaviour for short times, the general drawback of bivariate diffusion models such as Heston.

\subsection{The rough Bergomi model}
\begin{sidewaysfigure}[p] 
	\begin{subfigure}[b]{0.5\linewidth}
		\centering
		\includegraphics[width=\linewidth]{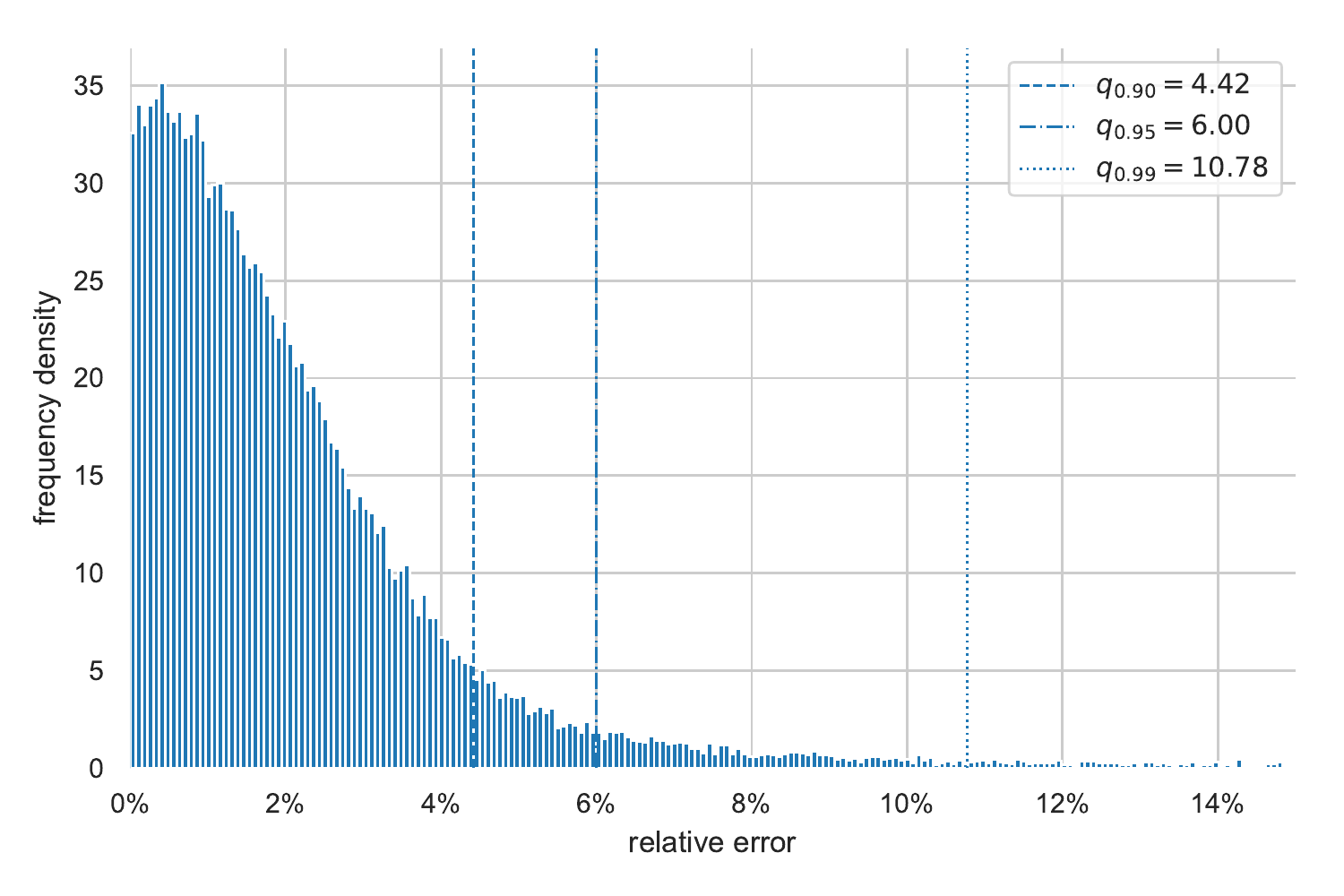} 
		\caption{Normalized histogram of RE \eqref{def:rel_err_iv} on testset with quantiles.} 
		\label{fig:rb_comp_graphs_c} 
	\end{subfigure}
	\begin{subfigure}[b]{0.5\linewidth}
		\centering
		\includegraphics[width=\linewidth]{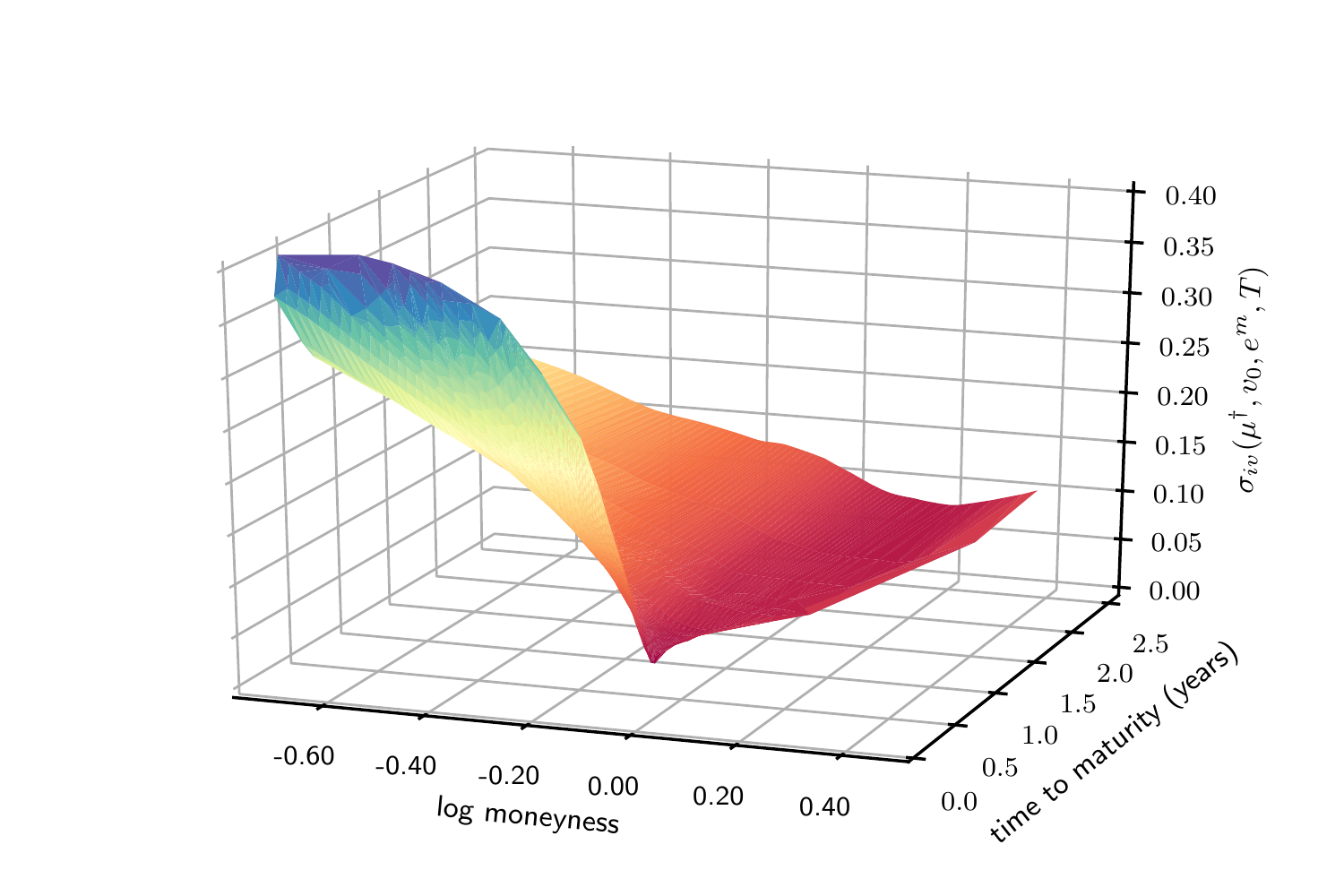} 
		\caption{Rough Bergomi IV surface computed using learned IV map $\varphi_\textrm{NN}$.} 
		\label{fig:rb_comp_graphs_a} 
	\end{subfigure}
	\begin{subfigure}[b]{0.5\linewidth}
		\centering
		\includegraphics[width=\linewidth]{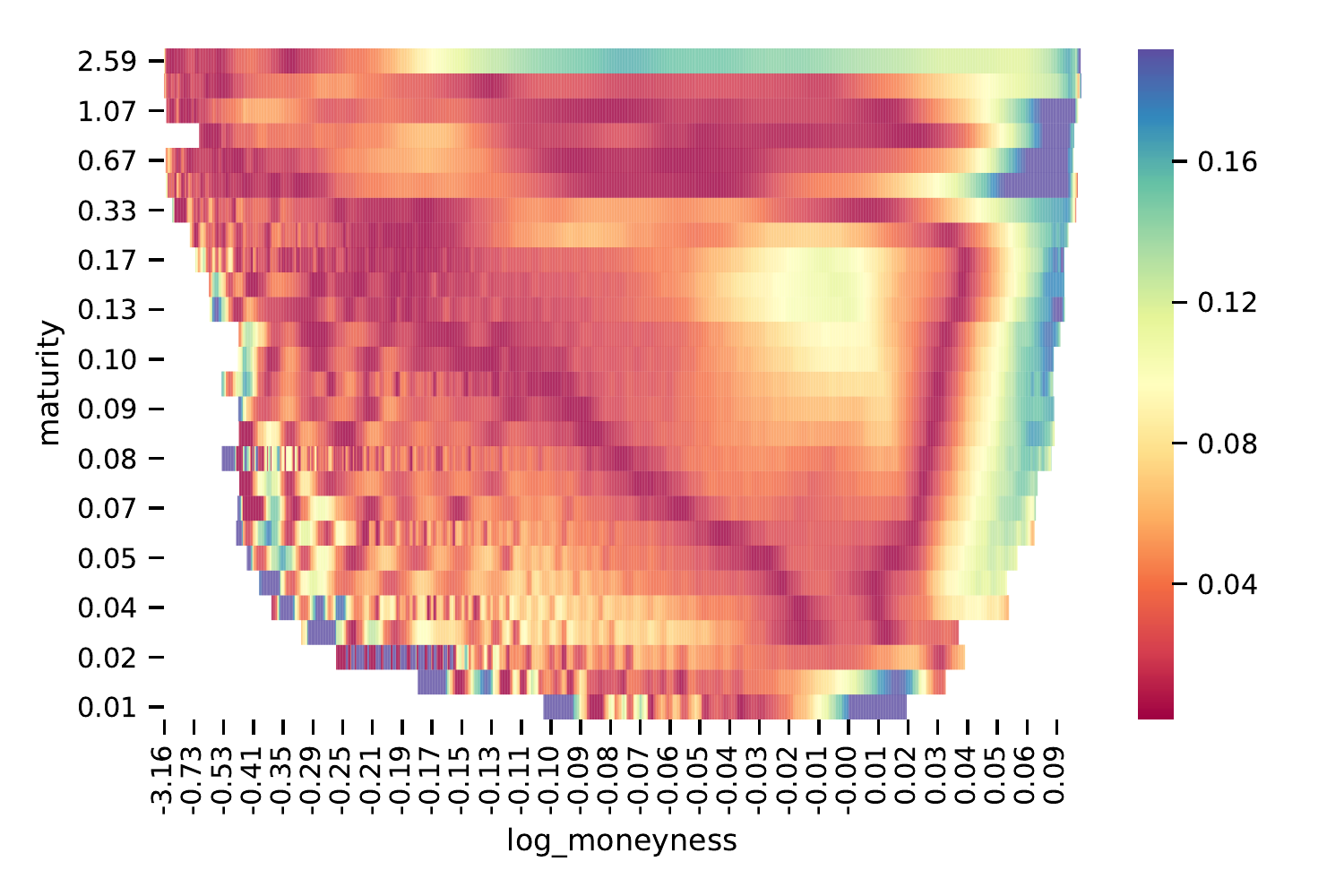} 
		\caption{Interpolated heatmap of $\textrm{RE}(\mu^\dagger, m, T)$ \eqref{def:rel_err_iv} for varying $m,T$.} 
		\label{fig:rb_comp_graphs_b} 
	\end{subfigure}
	\begin{subfigure}[b]{0.5\linewidth}
		\centering
		\includegraphics[width=\linewidth]{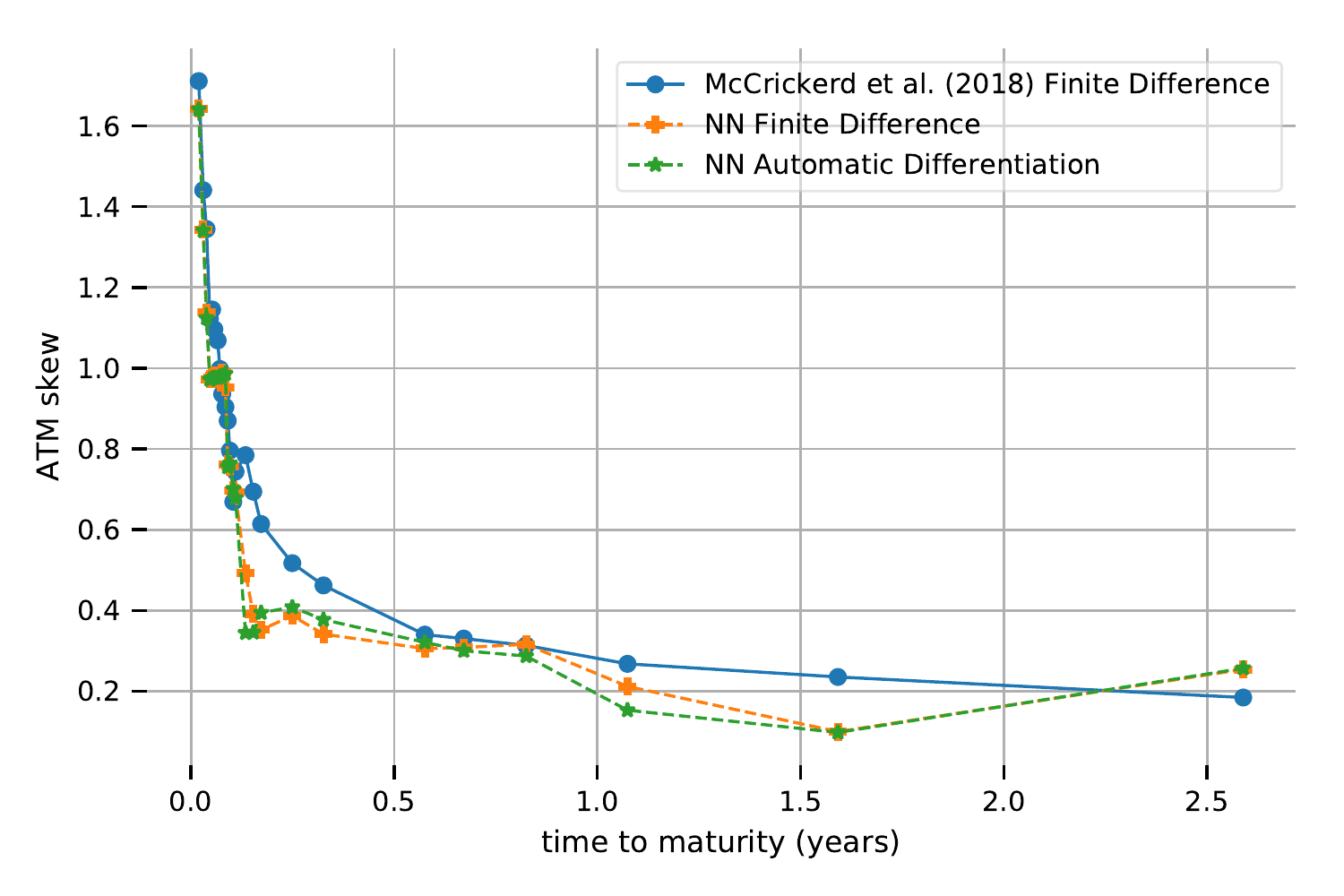} 
		\caption{Approximations to the ATM volatility skew.} 
		\label{fig:rb_comp_graphs_d} 
	\end{subfigure} 
	\caption{\textbf{Accuracy of the IV map $\varphinn$ as learned by the rough Bergomi ReLU FCNN.}}
\end{sidewaysfigure}
For simplicity, we consider the rough Bergomi model as introduced in Example \ref{ex:dc_roughbergomi} with a flat forward variance curve $\xi_0(t) = v_0 \in \mathbb{R}_+$ for $t\geq 0$. For the remainder of this work, we shall consider $v_0$ an additional model parameter. Again, following the approach outlined in Section \ref{sec:generation_synthetic_labdata}, we estimate $\mathcal{K}_{(M,T)}$ using SPX Option Price data, this time from 19th May 2017\footnote{ Thanks to Jim Gatheral for providing us with this data set.}. We do not restrict the option parameter region considered and learn the whole surface with parameter bounds given by 
$- 3.163 \leq  m \leq 0.391$ and $0.008 \leq T \leq 2.589$.
Of the one million synthetic data pairs sampled, 90\% are allocated to the training set and 5\% to validation and test sets respectively.

Recall that in this experiment we use the same network topology as in the Heston example. As is to be expected, the speed of single evaluations of the learned rough Bergomi IV map $\varphi_\textrm{NN}:\mathcal{I} \to \mathbb{R}_+$ and the associated Jacobian $\bm{J}_\textrm{NN}:~O~\to~\mathbb{R}^{N \times m}$ are hence of the same order with about $36$ms to compute both objects together, beating state of the art methods by magnitudes. Intuitively, the non-Markovian nature of rough Bergomi manifests itself in an increased model complexity and so it is unsurprising that the general accuracy of the rough Bergomi IV map $\varphinn$ on the rough Bergomi test set is lower than its counterpart on the Heston test set (cf. \ref{fig:rb_comp_graphs_c}). On the other hand, for fixed model parameters $\mu=\mu^\dagger$ (cf. Table \ref{ta:ref_parameter}), the implied volatility map $\varphinn$ recovers the characteristic rough Bergomi model IV surface (Figure \ref{fig:rb_comp_graphs_a}) with low relative error across most of the liquid parts of the IV surface (Figure \ref{fig:rb_comp_graphs_b}). It also exhibits the striking power law behaviour of the ATM volatility skew near zero (Figure \ref{fig:rb_comp_graphs_d}).

On the contrary, measuring the accuracy of the neural-network enhanced Levenberg-Marquardt scheme introduced in Section \ref{subsec:deep_cal} is not a straightforward task. To see why, consider the small-time asymptotic formula for the BS implied volatility $\sigma_\textrm{iv}$ of rough stochastic volatility models as derived by \citeA{BFGHS17}. With scaling parameter $\beta < \frac{2}{3} H$, their expansion applied to our setting yields
\begin{align}
\label{eq:bfghs_bsiv_asymptotic}
	\sigma_\textrm{iv}(e^{k_t}, t) = \sqrt{v_0} + \frac{1}{2} \rho \eta \; C(H) kt^\beta  + \mathcal{O}(t)
\end{align}
for small times $t \to 0$, time-scaled log moneyness $k_t = kt^{\frac{1}{2} - H + \beta}$ and constant $C(H)$ depending on $H$. Hence, at least for small times, all three model parameters enter multiplicatively either directly ($\rho$ and $\eta$) or indirectly ($H$)  into the second term in \eqref{eq:bfghs_bsiv_asymptotic} which corrects the crude estimate given by spot volatility. A decrease in $|\rho|$ could hence for example be offset by an adequate increase in $\eta$ and still yield the same IV. Mathematically speaking, for fixed moneyness and time to maturity, it is thus to be expected that the map $\varphinn$ is non-injective in its model parameters on large parts of its model parameter input domain. Quantifying the accuracy of the deep calibration scheme by computing any form of distance between true and calibrated model parameters in model parameter space is hence nonsensical.

\subsubsection{Bayesian parameter inference}
\begin{sidewaysfigure}[p]
	\begin{subfigure}[b]{0.5\linewidth}
		\begin{center}
			\includegraphics[width=\linewidth]{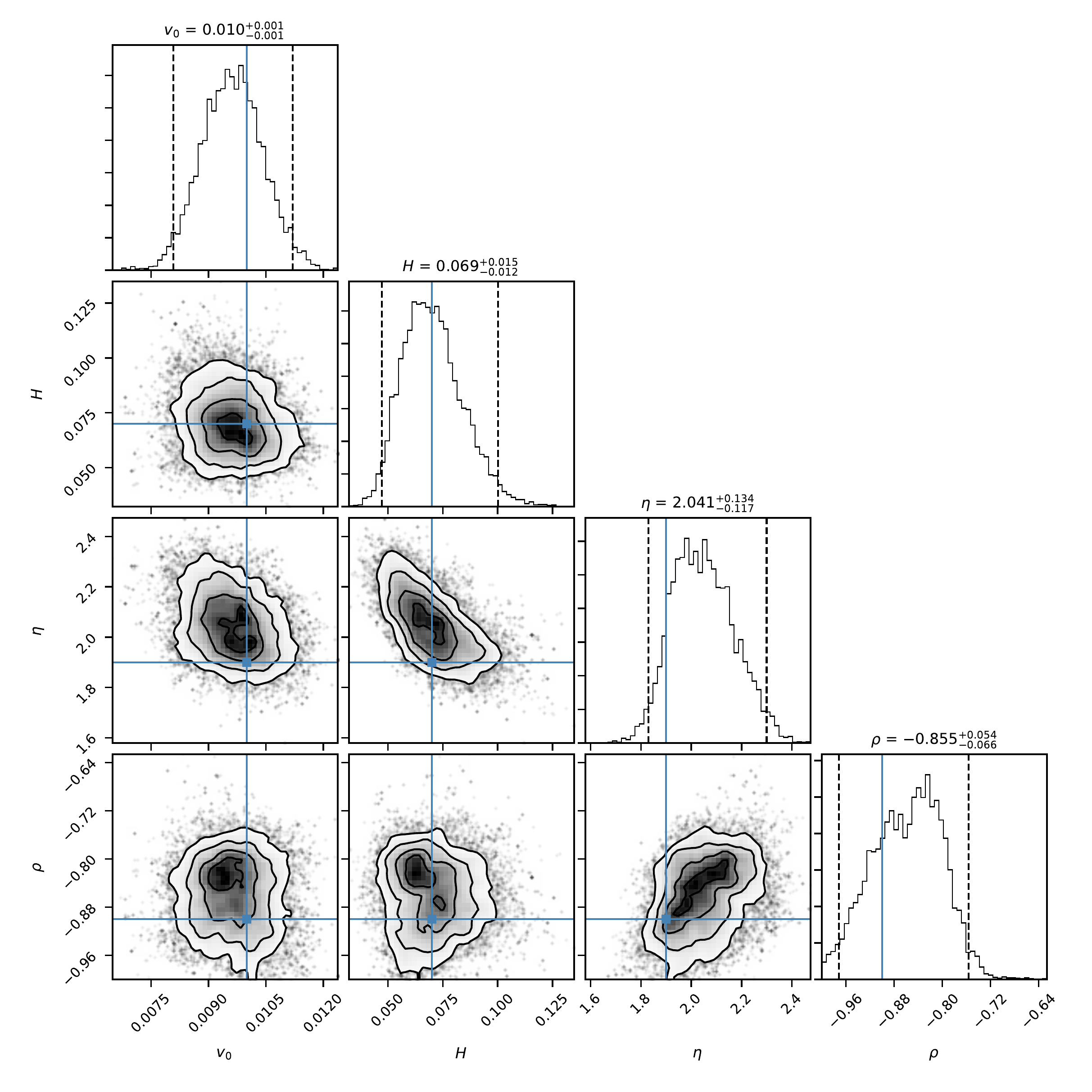}
		\end{center}
		\caption{Bayes calibration against synthetic IV surface computed for model parameters $\mu^\dagger$. Solid vertical blue lines indicate true parameter values.}
		\label{fig:rb_bayes_calibration_synthetic}
	\end{subfigure}\;
	\begin{subfigure}[b]{0.5\linewidth}
		\begin{center}
			\includegraphics[width=\linewidth]{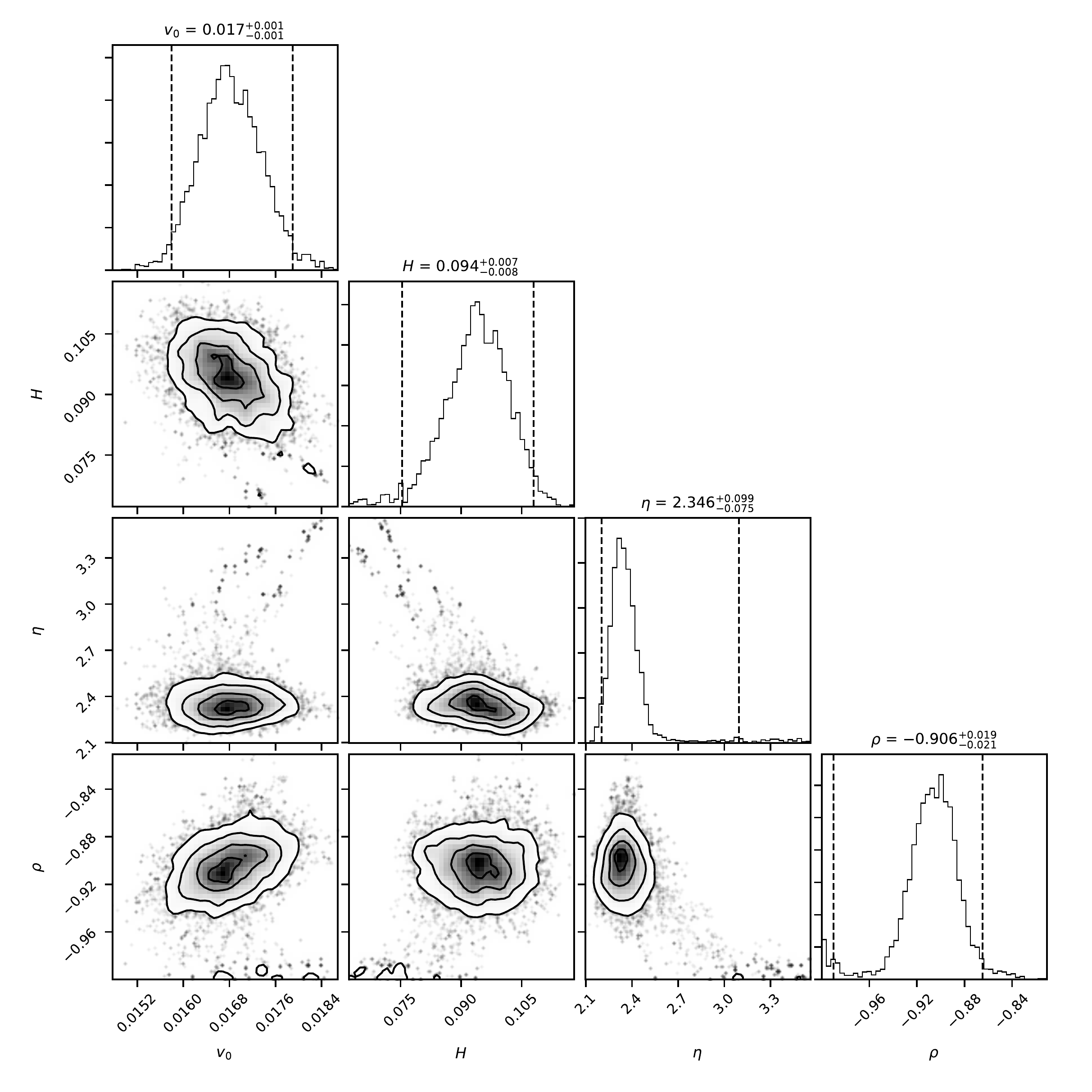}
		\caption{Liquidity-weighted Bayes calibration against SPX market IV surface from 19th May 2017. Liquidity proxies given by inverse bid-ask-spreads.}
				\label{fig:rb_bayes_calibration_marketdata}
		\end{center}
	\end{subfigure}
	\caption{\textbf{1d- and 2d-projections of 4d Bayesian posterior over rBergomi model parameters.} On diagonal, univariate histograms of model parameters with titles stating median $\pm$ the delta to $2.5\%$ and $97.5\%$ quantiles. Dashed vertical lines indicate those quantiles. Off-diagonal, 2d histograms of 2d projections of MCMC samples together with isocontours from 2d Gaussian KDE.}
\end{sidewaysfigure}

Intuitively, we are interested in \emph{quantifying the uncertainty} about model parameter estimates obtained by calibrating with the approximative IV map $\varphinn$. To this end, we switch to a Bayesian viewpoint and treat model parameters $\mu$ as random variables. The fundamental idea behind Bayesian parameter inference is to update prior beliefs $p(\mu)$ formalised in \eqref{eq:rb_prior} with the likelihood $p(\bm{y} \mid \mu)$ of observing a given IV point cloud $\bm{y}\in \mathbb{R}^N$ to deduce a posterior (joint) distribution $p(\mu \mid \bm{y})$ over model parameters $\mu$. 

Formally, for pairs $\left(M^{(i)}, T^{(i)}\right)$ of moneyness \& time to maturity, let an IV point cloud to calibrate against be given by $$\bm{y} = \left[y_1\left(M^{(1)}, T^{(1)}\right), \ldots, y_N\left(M^{(N)}, T^{(N)}\right)\right]^T \in \mathbb{R}^N$$
and analogously, collect model IVs for model parameters $\mu$ as follows
$$\bm{\varphinn}\left(\mu\right) = \left[\varphinn\left(\mu, M^{(1)}, T^{(1)}\right), \ldots, \varphinn\left(\mu, M^{(N)}, T^{(N)}\right)\right]^T \in \mathbb{R}^N.$$
We perform a liquidity-weighted nonlinear Bayes regression. Mathematically, for heteroskedastic sample errors $\sigma_i>0, i=1, \ldots, N$, we postulate 
\begin{align*}
	\bm{y} = \bm{\varphinn}\left(\mu\right) + \bm{\varepsilon}, \quad \bm{\varepsilon} \sim \mathcal{N}\left(0, \textrm{diag}[\sigma_1^2, \ldots, \sigma_N^2] \right)
\end{align*}
so that for some diagonal weight matrix $\bm{W} = \textrm{diag}\left[w_1, \ldots, w_N\right] \in \mathbb{R}^{N \times N}$, the liquidity-weighted residuals are distributed as follows
\begin{align*}
\bm{W}^{\frac{1}{2}}\left[\bm{y} - \bm{\varphinn}\left(\mu\right)\right] \sim  \mathcal{N}\left(0, \textrm{diag}[w_1\sigma_1^2, \ldots, w_N\sigma_N^2]\right) .
\end{align*} 
In other words, we assume that the joint likelihood $p\left( \bm{y} \mid \mu \right)$ of observing data $\bm{y}$ is given by a multivariate normal. In absence of an analytical expression for the posterior (joint) probability $p(\mu|\bm{y}) \propto p(\bm{y}|\mu) p(\mu)$, we approximate it numerically using MCMC techniques \cite{FHLG13} and plot the one- and two-dimensional projections of the four-dimensional posterior by means of an MCMC plotting library \cite{For16}.

We perform two experiments. First, fixing $\mu = \mu^\dagger$, we generate a synthetic IV point cloud 
$$\bm{y}_\textrm{synth}= \left[\varphitilde\left(\mu^\dagger, M^{(1)}, T^{(1)}\right), \ldots, \varphitilde\left(\mu^\dagger, M^{(N)}, T^{(N)}\right)\right] \in \mathbb{R}^N$$
using the reference method $\varphitilde$. Next, we perform a non-weighted Bayesian calibration against the synthetic surface and collect the numerical results in Figure \ref{fig:rb_bayes_calibration_synthetic}. If the map $\varphinn$ is sufficiently accurate for calibration, the computed posterior should attribute a large probability mass around $\mu^\dagger$. The results in Figure \ref{fig:rb_bayes_calibration_synthetic}  are quite striking in several ways: (1) From the univariate histograms on the diagonal it is clear that the calibration routine has identified sensible model parameter regions covering the true values. (2) Histograms are unimodal and its peaks close or identical to the true parameters. (3) The isocontours of the 2d Gaussian KDE in the off-diagonal pair plots for $(\eta, H)$ and $(\eta, \rho)$ show exactly the behaviour expected from the reasoning in the last section: Since increases or decreases in one of $\eta, H$ or $\rho$ can be offset by adequate changes in the others with no impact on the calculated IV, the Bayes posterior cannot discriminate between such parameter configurations and places equal probability on both combinations. This can be seen by the diagonal elliptic probability level sets.

In a second experiment, we want to check whether the inaccuracy of $\varphinn$ allows for a successful calibration against market data. To this end, we perform a liquidity-weighted Bayesian regression against SPX IVs from 19th May 2017.
For bid and ask IVs $a_i > 0$ and $b_i >0$ respectively, we proxy the IV of the mid price by $m_i :=  \frac{a_i + b_i}{2}$.
With spread defined by $s_i = a_i - b_i \geq 0$, all options with $s_i/m_i \geq 5\%$ are removed because of too little liquidity. Weights are chosen to be $w_i = \frac{m_i}{a_i - m_i} \geq 0$, effectively taking inverse bid-ask spreads as a proxy for liquidity. Finally, $\sigma_i$ are proxied by a fractional of the spread $s_i$. The numerical results in Figure \ref{fig:rb_bayes_calibration_marketdata} further confirm the accuracy of $\varphinn$: (1) As can be seen on the univariate histograms on the diagonal, the Bayes calibration has again identified sensible model parameter regions in line with what is to expected. (2) Said histograms are again unimodal with peaks at or close to values previously reported by \citeA{BFG16}. (3) Quite strikingly, at a first glance, the effect of the diagonal probability level sets in the off-diagonal plots as documented in Figure \ref{fig:rb_bayes_calibration_synthetic} cannot be confirmed here. However, the scatter plots in the diagrams do reveal some remnants of that phenomenon.

\subsubsection*{Acknowledgments}
The authors thank Jim Gatheral and Peter K. Friz for some helpful discussions and suggestions and acknowledge financial support through DFG Research Grants BA5484/1 and FR2943/2.

\bibliographystyle{apacite}

\end{document}